\documentstyle[12pt]{article}

\newcommand{\be}{\begin{equation}}
\newcommand{\ee}{\end{equation}}
\newcommand{\dlt}{\delta}
\newcommand{\Dlt}{\Delta}
\newcommand{\ra}{\rightarrow}
\newcommand{\vp}{\varphi}
\newcommand{\bt}{\beta}
\newcommand{\al}{\alpha}
\newcommand{\prt}{\partial}
\newcommand{\Om}{\Omega}
\newcommand{\om}{\omega}

\newcommand{\lbd}{\lambda}
\newcommand{\gm}{\gamma}
\newcommand{\Gm}{\Gamma}
\newcommand{\sgm}{\sigma}
\newcommand{\dgr}{\dagger}
\newcommand{\ep}{\varepsilon}
\newcommand{\bve}{{\bf e}}
\newcommand{\bn}{{\bf n}}
\newcommand{\br}{{\bf r}}
\newcommand{\bB}{{\bf B}}
\newcommand{\bS}{{\bf S}}
\newcommand{\bI}{{\bf I}}
\newcommand{\bk}{{\bf k}}
\newcommand{\bj}{{\bf j}}
\newcommand{\bA}{{\bf A}}

\begin{document}

\begin{center}
{\Large{\bf Coherent Nuclear Radiation } \\ [5mm]

V.I. Yukalov$^{1,2}$ and E.P. Yukalova$^3$} \\ [5mm]

{\it
$^1$Bogolubov Laboratory of Theoretical Physics \\
Joint Institute for Nuclear Research, Dubna 141980, Russia\\ [3mm]

$^2$Fachbereich Physik, Universitat Konstanz \\
Universitatsstra\ss e 10, Postfach 5560 M 674 \\
D-78434 Konstanz, Germany \\ [3mm]

$^3$Department of Computational Physics \\
Laboratory of Information Technologies \\
Joint Institute for Nuclear Research, Dubna 141980, Russia}

\end{center}

\vskip 2cm

\begin{abstract}

The main part of this review is devoted to the comprehensive description of
coherent radiation by nuclear spins. The theory of nuclear spin superradiance
is developed and the experimental observations of this phenomenon are
considered. The intriguing problem of how coherence develops from initially
incoherent quantum fluctuations is analysed. All main types of coherent
radiation by nuclear spins are discussed, which are: free nuclear induction,
collective induction, maser generation, pure superradiance, triggered
superradiance, pulsing superradiance, punctuated superradiance, and induced
emission. The influence of electron-nuclear hyperfine interactions and
the role of magnetic anisotropy are studied. Conditions for realizing spin
superradiance by magnetic molecules are investigated. The possibility of
nuclear matter lasing, accompanied by pion or dibaryon radiation, is briefly
touched.

\end{abstract}

\newpage

\section{Introduction}

Nuclei can radiate in different ways. For example, they emit gamma radiation
in the process of changing internal quantum states. An ensemble of nuclei,
emitting such hard photons, could form a coherent source, called gamma-ray
laser or just gamma laser. However, the problem of gamma lasers remains a
challenging but frustrating field of research, with not a great progress
in theory and practically no experimental achievements [1--3].

Contrary to this, there exists a type of coherent nuclear radiation that is
well documented experimentally and for which a detailed microscopic theory
has recently been developed. This is nuclear spin radiation. The main part
of the present review is just devoted to the phenomenon of coherent radiation
by nuclear spins. We, first, explain in simple words the physics of this
effect and survey the basic experiments where it was observed. Then we pass
to developing a comprehensive microscopic theory of strong nonlinear dynamics
of nuclear spins. A special attention is paid to a very intriguing problem of
how coherence arises from an incoherent quantum motion of randomly fluctuating
spins. We describe all principal regimes of spin superradiance and study the
influence of the hyperfine electron-nuclear coupling. We analyse the role of
the single-ion magnetic anisotropy that is so important in the spin radiation
of magnetic molecules. Finally, we note that not only photons can be emitted
coherently by nuclei. Thus, excited nuclear matter can produce coherent
radiation of gluons, pions, and dibaryons. Different types of coherent nuclear
radiation can find a variety of applications, some of which are discussed in
the review. The content of the latter is as follows:

\begin{enumerate}
\item
Introduction

\item
Nuclear Spin Superradiance

\item
Ensemble of Nuclear Spins

\item
Nuclear Spin Waves

\item
Scale Separation Approach

\item
Incoherent Quantum Stage

6.1. Radiation by Magnetic Dipoles

6.2. Resonator Nyquist Noise

6.3. Local Spin Fluctuations

\item
Regimes of Coherent Radiation

7.1. Transient Spin Superradiance

7.2. Pulsing Spin Superradiance

7.3. Induced Coherent Emission

\item
Electron-Nuclear Hyperfine Coupling

\item
Enhanced Nuclear Radiation

\item
Superradiance by Magnetic Molecules

\item
Pion and Dibaryon Radiation

\item
Conclusion

\end{enumerate}

\section{Nuclear Spin Superradiance}

Atomic systems, radiating at optical frequencies, exhibit a number of coherent
effects [4,5]. The majority of the latter have their counterparts in spin
systems, generating radiation at radio-frequencies. One of the most
interesting coherent effects is superradiance that may occur in both atomic
and spin systems. The possibility of superradiance in atomic systems was
predicted by Dicke [6] and the feasibility of organizing coherent motion of
spins was discussed by Bloembergen and Pound [7]. The modern status of nuclear
spin superradiance is presented in review [8].

By definition, superradiance is the process of {\it coherent spontaneous
radiation}. Being spontaneous, it is assumed to be self-organized. In general,
a coherent motion of spins can be realized by means of a strong external
magnetic field. But this would result in nuclear induction, which, though is
a coherent process, however is not superradiance. To realize the latter, one,
first, needs to prepare the spin ensemble in a nonequilibrium state. For this,
it is possible to polarize spins in one direction and after that to place
them in an external magnetic field of opposite direction. Such a system of
spins would be analogous to an ensemble of inverted atoms. Several other
similarities between atomic and spin systems are discussed in reviews [8,9].
Two important problems related to spin assemblies are: What is the cause
provoking spins to start their initial motion, after they have been inverted?
And what is the mechanism collectivizing the following spin motion, making
the latter coherent? The answer to the second question was proposed by
Bloembergen and Pound [7] who suggested to place the spin sample into an
electric coil being a part of an electric circuit with a natural frequency
tuned close to the Zeeman transition frequency of spins. This coupling with
a resonant electric circuit would produce a feedback field collectivizing the
spin motion. The answer to the first question, what is the cause starting the
spin motion, has not been understood for many years. This puzzle was solved
recently [10] and will be considered in detail in the following sections.
In brief, there are quantum fluctuations of spins that trigger the initial
spin motion.

One more question would be how to measure the radiation generated by moving
spins? The magnetodipole radiation, resulting from this motion, even being
completely coherent, would have the intensity of order
$$
I(t) \sim \frac{2}{3c^3}\; (\mu_0 \; I)^2\; \om_0^4\; N^2 \; ,
$$
where $c$ is the light velocity; $\mu_0\equiv\hbar\gm_n$, with $\gm_n$ being
the nucleus gyromagnetic ratio; $I$ is the nuclear spin; $\om_0$ is the
Zeeman transition frequency; $N$ is the total number of spins taking part
in the coherent radiation. Accepting the typical values $\mu_0\sim 10^{-23}$
erg/G, $\om_0\sim 10^8$ Hz, and $N\sim 10^{23}$, we would have $I(t)\sim
10^{-6}$ W, where $1$ W$=10^7$ erg/s is a watt. This is a rather small
quantity of the intensity $I(t)$, which would be different to notice. But this
radiation induces electric current in the coil, with the power of current
$P(t)$ being essentially larger than the intensity $I(t)$. Roughly speaking,
$$
\frac{P(t)}{I(t)} \sim \frac{\lbd^3}{V_c} \; ,
$$
where $\lbd$ is the radiation wavelength and $V_c$ is the coil volume. This
ratio, for $\lbd\sim 10^2-10^3$ cm and $V_c\sim 10$ cm$^3$, is of order
$10^5-10^8$, which makes $P(t)$ an easily measurable quantity. Thus, the
magnetodipole radiation of a spin sample is quite weak and practically
does not propagate into free space, but mainly is taken up by a resonant
coil surrounding the sample [11].

One should not confuse superradiance with other types of coherent radiation.
By definition, superradiance is {\it spontaneous collective emission}. The
term {\it spontaneous} implies that the process is {\it self-organized} but
not induced by external fields. And the word {\it collective} means that the
radiation characteristics are essentially influenced by collective effects.
For instance, the radiation intensity of $N$ radiators is proportional to
$N^\al$, with $\al>1$, while the radiated pulse is short, with the radiation
time proportional to $N^{1-\al}$. Depending on the relation between
characteristic times, there can arise different kinds of coherent radiation.
Among these typical times, the most important are: The crossover time $t_c$,
separating the quantum incoherent stage of spin motion from their coherent
rotation; the pulse time $\tau_p$ of an emitted pulse; and the dephasing
time $T_2$. In the case, when there are no external transverse fields,
one can distinguish seven coherent radiation regimes:

\vskip 2mm

(1) {\it Free induction} ($t_c=0,\; \tau_p\geq T_2$).

\vskip 1mm

Coherence of radiation here is not self-organized but imposed upon the system
by an external field. This is the standard nuclear free decay induced by an
initial coherent pulse or, equivalently, by an essential transverse
polarization [12].

\vskip 2mm

(2) {\it Collective induction} ($t_c=0,\; \tau_p < T_2$).

\vskip 1mm

The shortening of the radiation damping is due to the coupling with a
resonator. However, the process is mainly governed not by a self-organization
but by a strong external coherent field or by an initial transverse polarization
[7].

\vskip 2mm

(3) {\it Maser generation} ($t_c>0,\; \tau_p\geq T_2$).

\vskip 1mm

There are no external coherent fields, though some incoherent nonresonant
pumping may exist. Self-organization becomes crucial. But the emitted pulses
are not sufficiently narrow, implying that self-organization is not yet high.
Although this is a spontaneous coherent process, it is not yet a genuine
superradiance. Such a type of coherent maser generation was observed in many
experiments [13--16]. The physics of this process is analogous to the known
laser generation [17--19].

\vskip 2mm

(4) {\it Pure superradiance} ($t_c>0,\; \tau_p<T_2$).

\vskip 1mm

This is a purely self-organized process, developing from an incoherent chaotic
stage to a highly correlated spin motion. Though some of the features of spin
superradiance are similar to those of atomic superradiance [4,20--22], there
are also several principal differences. Experiments on and theory of spin
superradiance will be described in detail below.

\vskip 2mm

(5) {\it Triggered superradiance} ($t_c>0,\; \tau_p<T_2$).

\vskip 1mm

In this process, self-organization is crucially important, but there is as
well a weak external resonant field triggering and influencing the behaviour
of spins. There are similarities with triggered superradiance in optics [20].

\vskip 2mm

(6) {\it Pulsing superradiance} ($t_c>0,\; \tau_p<T_2$).

\vskip 1mm

This regime differs from pure superradiance by the existence of a series of
superradiant pulses, instead of a single superradiant burst. To realize such
a pulsing superradiance, it is necessary to maintain a sufficiently high
level of spin polarization by means of an incoherent pumping. The latter
can be done, e.g., by dynamic nuclear polarization [12,23,24].

\vskip 2mm

(7) {\it Punctuated superradiance} ($t_c>0,\; \tau_p<T_2$).

\vskip 1mm

Here, similar to pulsing superradiance, there is a series of superradiant
pulses; however this is achieved not by supporting nuclear polarization but
by means of resonant external fields or forces. All regimes of spin
superradiance will be thoroughly considered in the following sections.

\vskip 2mm

From the above classification it follows that superradiance, in addition
to being a coherent and spontaneous emission, should correspond to short
pulses with a finite crossover time, so that
$$
t_c>0\; , \qquad \tau_p < T_2 \; .
$$
Thus, the correct and full definition of superradiance would be:

\vskip 1mm

{\it Superradiance is a coherent spontaneous emission by an ensemble of
radiators of short pulses peaked at a time that is larger than the crossover
time, with the duration of each pulse shorter than the dephasing
time}.

\vskip 2mm

This definition suits for any kind of superradiance, whether it is atomic
or spin superradiance. In what follows, we concentrate on spin superradiance
produced by an ensemble of nuclear spins. Initially, we shall give a brief
survey of experiments observing nuclear spin superradiance.

{\it Pure spin superradiance} by nuclei was first observed in Dubna [25,26]
and confirmed in St. Petersburg [27,28]. These experiments were accomplished
with propanediol C$_3$H$_8$O$_2$. This is a material rich of protons, with
the density $\rho_H\approx 4\times 10^{22}$ cm$^{-3}$. The proton spins were
polarized, by dynamic nuclear polarization, in an external magnetic field
$B_0\sim 1$ T, which corresponds to the Zeeman frequency $\om_0\sim 10^8$ Hz.
The sample was refrigerated to low temperatures $T\sim 0.1$ K, at which the
nuclear spin-lattice relaxation was strongly suppressed, with the longitudinal
relaxation time $T_1\sim 10^5$ s. The transverse dephasing time, due to dipole
spin-spin interactions, was $T_2\sim 10^{-5}$ s. The coupled resonant electric
circuit had a quality factor $Q\sim 100$ and a ringing time $\tau\sim
10^{-6}$ s.

Similar experiments were accomplished in Bonn [29] with butanol C$_4$H$_9$OH
and ammonia NH$_3$. In these materials, the proton density $\rho_H\sim
10^{23}$ cm$^{-3}$. The characteristic Zeeman frequency was $\om_0\sim 10^8$
Hz. The materials were kept at low temperature, when the spin-lattice
relaxation is suppressed, with the longitudinal relaxation time $T_1\sim
10^5$ s. The transverse dephasing time was $T_2\sim 10^{-5}$ s. The quality
factor of the resonator electric circuit was $Q\sim 30$, and the ringing
time $\tau\sim 5\times 10^{-7}$ s.

Such proton-rich materials are widely used as targets in studying the
scattering of particle beams from accelerators. It would, of course, be
possible to involve for experiments on spin superradiance other materials
employed as targets in high-energy scattering studies. For instance, a good
candidate would be pentanol C$_5$H$_{12}$O, whose proton density is $\rho_H
\approx 6\times 10^{22}$ cm $^{-3}$.

{\it Pulsing spin superradiance} was explored in Z\"urich [30--33] on the
basis of nuclei of $^{27}$Al inside the ruby crystal Al$_2$O$_3$, where the
density $\rho_{Al}\approx 4\times 10^{22}$ cm$^{-3}$. The nuclei $^{27}$Al
have spins $I=5/2$. The crystal was oriented in an external magnetic field
$B_0\sim 1$ T so that the fully resolved structure of the five $\Dlt m=\pm 1$
transition lines could be clearly seen. When a resonant circuit is tuned to
a selected transition line, then $^{27}$Al spins form an effective two-level
system. In experiments [30--33], the circuit was usually tuned to the central
$\{-\frac{1}{2},\frac{1}{2}\}$ line, with a transition frequency
$\om_0\sim 10^8$ Hz. At low temperatures around $T\sim 1$ K, the spin-lattice
relaxation time was $T_1\sim 10^5$ s and the transverse dephasing time was
$T_2\sim 10^{-5}$ s. The resonant circuit had the quality factor $Q\sim 100$
and ringing time $\tau\sim 10^{-6}$ s. The inversion of spin polarization
was permanently supported by means of dynamic nuclear polarization with the
pumping rate $\Gm_1^*$ between $0.01$ and $10$ s$^{-1}$.

A typical experimental setup employed for observing spin superradiance [25,26]
is shown in Fig. 1. The detected superradiance pulse has the form illustrated
in Fig. 2 for two different initial spin polarizations. The higher is the
prepared inversion, the stronger is the superradiant burst.

The influence of the passive resonant circuit, coupled to a nuclear
spin system, was also studied in nuclear spin echo experiments
[34--36]. In the ferrite Li$_{0.5}$Fe$_{2.5}$O$_4$, the spins of
nuclei $^{57}$Fe were tuned to a resonant circuit with a frequency
about $7\times 10^7$ Hz. In the compound Co$_2$MnSi, the working
substance was the nuclei $^{59}$Co, whose spins were coupled to
a circuit of frequency $1.4\times 10^8$ Hz. The experiments
[34--36] showed that the presence of the resonant circuit can
enhance the intensity of the nuclear spin echo signal by up to
40$\%$, as compared to the case without coupling to such a resonant
circuit [37].

The theoretical description of spin dynamics in the presence of a
resonant circuit has been commonly done on the basis of the Bloch plus
Kirchhoff equations [27--33,38--40]. Analytical solutions were usually
obtained by involving the adiabatic approximation. However, the latter
approximation is only valid for dynamics close to stationary and it
does not correctly describe transient effects, such as superradiant
bursts. What is more, the Bloch equations presuppose the existence
of coherent motion from the beginning, since these are classical
equations. Such equations principally are not able to depict
self-organized coherent phenomena, as pure superradiance, which
develops from an initially incoherent quantum stage. This principal
defect of Bloch equations cannot be overcome even if one employs more
elaborate approximations [41,42] based on the averaging technique
[43]. As was first shown in Ref. [10], for the correct description
of pure spin superradiance it is crucially important to take account
of stochastic local spin fluctuations.  This conclusion was also
confirmed by numerical calculations [44--46].

Computer simulations of spin superradiance has been accomplished
[47--51], investigating different regimes of coherent spin motion.
A typical result of simulations is shown in Fig. 3. The weak points
of such simulations are: First, one has to deal with a rather
limited number of spins, say $10^2$ or $10^3$, because of which
the obtained solutions give only a qualitative picture. Second,
in such simulations, spins are treated as classical vectors, hence,
quantum fluctuations, so important at the initial stage of motion,
are not properly considered. Third, numerical results do not always
present clear physical explanations of the studied processes. This
is especially so, when there are tens of parameters to be varied,
while in reality only some combinations of these parameters are
important.

To give a thorough description of nuclear spin superradiance, it
is necessary to develop a good microscopic theory of spin dynamics
in a sample coupled to a resonant electric circuit. The feedback
field created by the resonator is crucial for organizing the coherent
motion of spins, moving from a strongly nonequilibrium state. Note
that for spin motion close to equilibrium, as in nuclear magnetic
resonance experiments, the feedback field is not of much significance
[52]. For describing such a purely self-organized process as pure
spin superradiance, it is vital to take into account local quantum
spin fluctuations, which trigger the initial spin motion and define
the delay time of the superradiant burst [10].

A comprehensive microscopic theory of spin superradiance, based
on realistic spin Hamiltonians [12,23,53,54] has been developed
[10,55--60]. This theory for the first time has made it possible
to discover the actual origin of pure spin superradiance [10,55--57],
to give an accurate description of various regimes of nonlinear spin
dynamics [55--59], to present an explicit picture of how coherence
develops from chaotic spin fluctuations [10,60], to consider spin
superradiance in different materials, such as polarized nuclear
paramagnets [10,56--59], ferromagnets, ferrimagnets [61--63], and
molecular magnets [60], to analyse the superradiant operation of
spin masers [60,64,65], and to advance the feasibility of punctuated
spin superradiance [66]. Brief survey of this theory was done in
reviews [8,9]. In the following sections, we shall give a detailed
account of the theory and of several its applications.

\section{Ensemble of Nuclear Spins}

Consider a solid sample containing $N$ nuclear spins $I_i$ enumerated
by the index $i=1,2,\ldots,N$. The Hamiltonian of the nuclear system
can be written as
\be
\label{1}
\hat H =\sum_i \hat H_i + \frac{1}{2}\;
\sum_{i\neq j} \hat H_{ij} \; ,
\ee
with $\hat H_i$ being related to individual spins and $\hat H_{ij}$
corresponding to spin interactions. The single-spin term
\be
\label{2}
\hat H_i = - \mu_0 \bB\cdot\bI_i
\ee
is the Zeeman energy, where $\mu_0\equiv g_I\mu_N=\hbar\gm_n$; with
$g_I$ being the Land\'e factor for a nucleus of spin $I$; $\mu_N$,
nuclear magneton; $\gm_n$, nuclear gyromagnetic ratio. Generally, for
nuclei, $\mu_0$ can be positive as well as negative. The total magnetic
field
\be
\label{3}
\bB =B_0\bve_z +(B_1+H)\bve_x
\ee
contains external longitudinal, $B_0$, and transverse, $B_1$, magnetic
fields, and also a feedback field $H$ of the resonant electric circuit.
The pair terms
\be
\label{4}
\hat H_{ij} = \sum_{\al\bt}\; D_{ij}^{\al\bt} I_i^\al\; I_j^\bt
\ee
in the Hamiltonian (1) correspond to dipolar spin interactions through
the dipolar tensor
\be
\label{5}
D_{ij}^{\al\bt}  = \frac{\mu_0^2}{r_{ij}^3}\;
\left ( \dlt_{\al\bt} - 3 n_{ij}^\al\; n_{ij}^\bt\right ) \; ,
\ee
in which $\al,\bt=x,y,z$ and
$$
r_{ij}\equiv |\br_{ij}|\; , \qquad \bn_{ij} \equiv
\frac{\br_{ij}}{r_{ij}} \; , \qquad \br_{ij} \equiv \br_i - \br_j \; .
$$
The dipolar tensor enjoys the properties
\be
\label{6}
\sum_{j(\neq i)} \; D_{ij}^{\al\bt} = 0 \; , \qquad
\sum_\al \; D_{ij}^{\al\al} = 0 \; ,
\ee
of which the second is exact and the first one is asymptotically
exact for a macroscopic (in all directions) sample with a large number
of spins $N\gg 1$.

The single-spin term (2), with the total magnetic field (3), can be
written as
\be
\label{7}
\hat H_i = -\mu_0 B_0 I_i^z - \; \frac{1}{2}\; \mu_0 \left ( B_1 +
H \right ) \left ( I_i^+ + I_i^- \right ) \; ,
\ee
where $I_j^\pm \equiv I_j^x \pm i I_j^y$ are the ladder spin operators.
Employing the latter, the interaction term (4) can be presented in the
form
\be
\label{8}
\hat H_{ij} = a_{ij} \left ( I_i^z\; I_j^z -\; \frac{1}{2}\; I_i^+
I_j^-\right ) + b_{ij}\; I_i^+\; I_j^+ + b^*_{ij}\; I_i^-\; I_j^-
+ 2c_{ij}\; I_i^+\; I_j^z + 2\; c_{ij}^*\; I_i^-\; I_j^z \; ,
\ee
in which the notation
\be
\label{9}
a_{ij} \equiv D_{ij}^{zz}\; , \qquad c_{ij}\equiv \frac{1}{2}
\left ( D_{ij}^{xz} - i D_{ij}^{yz} \right ) \; , \qquad
b_{ij} \equiv \frac{1}{4} \left ( D_{ij}^{xx} - D_{ij}^{yy} - 2i\;
D_{ij}^{xy} \right )
\ee
is introduced. From the first of Eqs. (6), it follows that
\be
\label{10}
\sum_{j(\neq i)} \; a_{ij} =\sum_{j(\neq i)} \; b_{ij} =
\sum_{j(\neq i)}\; c_{ij} = 0 \; .
\ee
Note also that the quantities (9) are symmetric, so that $a_{ij}=
a_{ji}$, $b_{ij}=b_{ji}$, and $c_{ij}=c_{ji}$.

In the total magnetic field (3), the external longitudinal and
transverse fields $B_0$ and $B_1$, respectively, are supposed to
be prescribed. The resonator feedback field $H$ is created by the
electric current of the coil surrounding the sample. The orientation
of the coordinate axes with respect to the latter is illustrated
in Fig. 4. The electric circuit is characterized by resistance $R$,
inductance $L$, and capacity $C$. The spin sample is inserted into
a coil of $n$ turns, length $l$, and cross-section area $A_c$. The
electric current in the circuit is described by the Kirchhoff
equation
\be
\label{11}
L\; \frac{dj}{dt} + Rj + \frac{1}{C} \;
\int_0^t \; j(t')\; dt' =  E_f -\; \frac{d\Phi}{dt} \; ,
\ee
in which $E_f$ is an electromotive force and
\be
\label{12}
\Phi = \frac{4\pi}{c}\; n\; A_c\; \eta\; m_x
\ee
is a magnetic flux formed by the $x$-component of the magnetization
density
\be
\label{13}
m_x = \frac{\mu_0}{V}\; \sum_i < I_i^x> \; ,
\ee
where $\eta\approx V/V_c$ is a filling factor, $V$ is the sample
volume, and the brackets $<\ldots>$ imply statistical averaging.

The electric current, circulating over the coil, creates a magnetic
field
\be
\label{14}
H = \frac{4\pi n}{c\;l} \; j \; .
\ee
Let us employ the standard notation for the circuit natural frequency
\be
\label{15}
\om \equiv \frac{1}{\sqrt{L C}} \; , \qquad L \equiv
4\pi\; \frac{n^2 A_c}{c^2\;l} \; ,
\ee
and the circuit damping
\be
\label{16}
\gm \equiv \frac{R}{2L} = \frac{\om}{2Q} \equiv \frac{1}{\tau} \; ,
\qquad Q \equiv \frac{\om L}{R} \; ,
\ee
where $Q$ is the resonator quality factor and $\tau$ is called the
circuit ringing time. Define the reduced electromotive force
\be
\label{17}
h \equiv \frac{c E_f}{nA_c\gm} \; .
\ee
Then the Kirchhoff equation (11) can be transformed to the equation
\be
\label{18}
\frac{dH}{dt} + 2\gm H + \om^2 \;
\int_0^t H(t')\; dt' = \gm h - 4\pi\eta\; \frac{dm_x}{dt}
\ee
for the feedback magnetic field created by the coil.

The feedback equation (18) can be presented in another equivalent
form that proved to be very convenient for solving the evolution
equations [55--60]. For this purpose, by involving the method of
Laplace transforms and employing the transfer function
\be
\label{19}
G(t) = \left ( \cos\om' t -\; \frac{\gm}{\om'} \;
\sin\om' t \right ) \; e^{-\gm t} \qquad \left ( \om' \equiv
\sqrt{\om^2-\gm^2} \right ) \; ,
\ee
we find the integral representation
\be
\label{20}
H = \int_0^t G(t -t') \left [ \gm h(t') - 4\pi \eta \dot{m}_x(t')
\right ] \; dt' \; ,
\ee
where the overdot means, as usual, time derivative.

\section{Nuclear Spin Waves}

In order to understand the nature of the mechanisms triggering the spin
motion, it is necessary to pay attention to the possibility of arising
nuclear spin waves. The latter are known to exist in ferromagnets, where
ferromagnetism is caused by electron spins [54]. To produce spin waves,
nuclear spins should be somehow ordered, at least locally. The dipolar
interactions (4), by themselves, usually are not able to order nuclear
spins. However, in the presence of a sufficiently strong external magnetic
field, typical of nuclear magnetic resonance experiments, some kind of the
intermediate range ordering may appear among nuclear spins. This mid-range
ordering can be observed by different NMR techniques [67].

The situation with nuclear spins, forming a paramagnetic matter, seems to
be analogous to what happens in paramagnetic systems of electron spins.
Therefore, it is useful to make a retrospective journey into the problem
of spin waves in electron paramagnetic assemblies. This may better explain
the physics of the similar effects in nuclear spin samples.

The theory of electron spin waves in nonmagnetic metals was developed
by Silin [68,69] on the basis of the Landau Fermi-liquid phenomenological
picture [70,71]. Good surveys of the Silin theory can be found in
[72--74]. Spin waves in nonmagnetic metals can exist only in the presence
of an external magnetic field, and do not exist when this field is absent.
To be well defined, spin waves require that the external magnetic field
be sufficiently strong, so that the spin resonance frequency be larger
than the spin-wave attenuation. Spin waves, in the field of about $10^3$
G, were observed in several paramagnetic metals [75--78], such as Na,
K, Rb, Cs, and Al. Calculation of electro-magnetic fields transmitted
through metallic films by spin waves involves the usage of boundary
conditions defined by the type of spin scattering on the surface of
metals. Several kinds of these conditions were employed in the theory
of electron spin resonance [79--87]. The Dyson boundary condition [79]
was used for studying the influence of the surface magnetic anisotropy
on the amplitudes of excited spin waves and on the transition coefficient
under spin-wave resonance [89--91]. Actually, experiments on spin resonance
cannot distinguish between different boundary conditions [85]. Therefore,
the usage of the simple Dyson condition for calculating the signals
transmitted by spin waves [89--91] was justified. The surface impedance
of a semi-bounded metal was expressed through the characteristics of
infinite matter [90]. The possibility of exciting spin waves in a
semi-bounded paramagnetic sample was advanced [92,93]. Thus, electron
spin waves do exist in paramagnetic metals, provided that a sufficiently
strong external magnetic field is applied. All this sets us thinking that
an external magnetic field could also ensure the appearance of nuclear
spin waves in paramagnetic nuclei assemblies.

Let us consider a paramagnetic system of nuclear spins, described by the
Hamiltonian  (1). The spin operators satisfy the Heisenberg equations of
motion
$$
i\hbar \frac{d I_i^\al}{dt} =\left [ I_i^\al,\hat H\right ]
$$
and the commutation relations
$$
\left [ I_i^+,I_j^-\right ] = 2\dlt_{ij}\; I_i^z \; , \qquad
\left [ I_i^z,I_j^\pm\right ] = \pm \dlt_{ij}\; I_i^\pm \; .
$$
To write down the evolution equations in an explicit way, it is
convenient to introduce the notation
$$
\xi_0 \equiv \frac{1}{\hbar} \; \sum_{j(\neq i)} \left ( a_{ij}\;
I_j^z + c^*_{ij}\; I_j^- + c_{ij}\; I_j^+\right ) \; ,
$$
\be
\label{21}
\xi \equiv \frac{i}{\hbar} \; \sum_{j(\neq i)} \left ( 2 c_{ij}\;
I_j^z  -\; \frac{1}{2}\; a_{ij}\; I_j^- + 2b_{ij}\; I_j^+ \right )
\ee
for the {\it local fields} acting on a spin from the side of other
spins. For short, we do not label the local fields (21) by an index
$i$. Also, define the {\it effective force}
\be
\label{22}
 f\equiv -\; \frac{i}{\hbar} \; \mu_0 (B_1 +H) + \xi \; .
\ee
Then the evolution equations for the spin operators can be presented
in the form
$$
\frac{dI_i^-}{dt} = i\left ( \frac{1}{\hbar}\; \mu_0 B_0 -
\xi_0\right ) I_i^- + f\; I_i^z \; ,
$$
\be
\label{23}
\frac{dI_i^z}{dt} = -\; \frac{1}{2} \left ( f^+\; I_i^-
+ I_i^+\; f \right ) \; .
\ee

Denoting the statistical average of an operator $I_i^\al$ as
$<I_i^\al>$, we may write the identity
\be
\label{24}
I_i^\al \equiv \; < I_i^\al>\; + \dlt\; I_i^\al \; , \qquad
\dlt\; I_i^\al \equiv I_i^\al -\; < I_i^\al> \; ,
\ee
where $\dlt I_i^\al$ describes an operator deviation from the average
value $<I_i^\al>$. Also, without the loss of generality, we may assume
that the averages
\be
\label{25}
<I_i^\al> \; = \; <I_j^\al>
\ee
do not depend on the indices $i,j$, thus, transferring this dependence to
the deviation term $\dlt I_i^\al$. Then, because of Eqs. (10), the local
fields (21) become
$$
\xi_0 = \frac{1}{\hbar} \; \sum_{j(\neq i)} \left ( a_{ij}\;
\dlt I_j^z + c^*_{ij}\; \dlt I_j^- + c_{ij}\; \dlt I_j^+\right ) \; ,
$$
\be
\label{26}
\xi  = \frac{i}{ \hbar} \; \sum_{j(\neq i)} \left ( 2 c_{ij}\;
\dlt I_j^z  -\; \frac{1}{2}\; a_{ij}\; \dlt I_j^- + 2b_{ij}\; \dlt
I_j^+ \right ) \; .
\ee
This better clarifies the meaning of the local fields $\xi_0=\dlt\xi_0$,
and $\xi=\dlt\xi$, actually, corresponding to {\it local field fluctuations}.

To show that an external magnetic field $B_0$ can support the appearance
of nuclear spin waves, let us consider a stationary situation, when
$B_0=const$ and $B_1=H=0$, so that there is a nonzero polarization
$<I_i^z>\not\equiv 0$, but $<I_i^\pm>=0$. In such a case, $I_i^\pm=\dlt
I_i^\pm$. Equating in Eqs. (23) the terms linear with respect to spin
deviations, we have
\be
\label{27}
\frac{d}{dt}\; I_i^- = \frac{i}{\hbar}\; \mu_0\; B_0\; I_i^-
+\; <I_i^z>\; \xi \; , \qquad \frac{d}{dt}\; \dlt\; I_i^z =0\; .
\ee
Since $\dlt I_i^z=const$, we may set, by accepting the zero initial
condition, that $\dlt I_i^z=0$. For the raising and lowering spin
operators, we employ the Fourier transforms
\be
\label{28}
I_j^\pm =\sum_k I_k^\pm\; \exp\left (\mp i\bk\cdot\br_j \right )\; ,
\qquad I_k^\pm = \frac{1}{N} \sum_j I_j^\pm \; \exp\left (
\pm i\bk\cdot \br_j\right ) \; .
\ee
Strictly speaking, the latter are exactly valid only if all spins
were located in sites of an ideal crystalline lattice. When the
studied spin sample does not form such an ideal lattice,
transformation (28) can be treated as approximate. In that case,
the summation over spins should be limited by a finite number of
them inside a volume, with an effective size $L^3_{eff}$, where
the matter can be regarded as approximately arranged in a lattice.
The length $L_{eff}$ is an effective linear size of uniformity.

Introduce the transforms
\be
\label{29}
a_k \equiv \sum_{j(\neq i)} a_{ij}\; \exp\left (-i\bk\cdot\br_{ij}
\right ) \; , \qquad
b_k \equiv \sum_{j(\neq i)} b_{ij}\; \exp\left (-i\bk\cdot\br_{ij}
\right ) \; .
\ee
Since $a_{ij}=a_{ji}$ and it is real, $a_k$ is also real. And, because
of Eqs. (10), $a_0=b_0=0$. Then, using Eqs. (28) and (29), we reduce
the first of Eqs. (27) to the equation
\be
\label{30}
\frac{d}{dt}\; I_k^- = - i\mu_k\; I_k^- + i\lbd_k\; I_k^+ \; ,
\ee
in which
\be
\label{31}
\mu_k \equiv \frac{a_k}{2\hbar}\; <I_i^z> \; - \; \frac{1}{\hbar}\;
\mu_0\; B_0 \; , \qquad \lbd_k \equiv \frac{2b_k}{\hbar}\;
< I_i^z>\; .
\ee
The solution to Eq. (30) can be presented as
\be
\label{32}
I_k^- = u_k \; e^{-i\om_k t} + v_k^*\; e^{i\om_k t} \; ,
\ee
with the spectrum of excitations
\be
\label{33}
\om_k =\sqrt{\mu_k^2 - |\lbd_k|^2} \; .
\ee
The uniform excitation, corresponding to $k=0$, yields
\be
\label{34}
\om_0 = \frac{1}{\hbar}\; |\mu_0\; B_0 | \; .
\ee
The procedure of defining the excitation spectrum, as is used above,
is analogous to that employed for ferromagnets [94].

The spectrum (33) can describe nuclear spin waves if it is positive.
The requirement $\om_k>0$ leads to the {\it stability condition}
\be
\label{35}
\left | 2\mu_0 B_0 -\; <I_i^z> a_k\right | > 4\left |
<I_i^z> b_k \right | \; .
\ee
From here, it is evident that in a paramagnet without a polarization,
when $<I_i^z>=0$, and without an external field, i.e. $B_0=0$, there
are no nuclear spin waves. But if there exists a polarization and
an external field, that is, $<I_i^z>\neq 0$ and $B_0\neq 0$, then
condition (35) can be valid for a sufficiently strong external field
$B_0$. Therefore, nuclear spin waves are possible in a paramagnet
placed in a magnetic field.

In the long-wave limit, the spectrum (33) is given by the equation
\be
\label{36}
\om_k^2 \simeq \om_0^2 +\; <I_i^z>\; \frac{\mu_0B_0}{2\hbar^2} \;
\sum_{j(\neq i)} a_{ij} \left ( \bk \cdot \br_{ij} \right )^2 \; ,
\ee
where $k\ra 0$. In the interval
\be
\label{37}
L_{eff}^{-1} \ll k \ll a^{-1} \; ,
\ee
where $L_{eff}$ is an effective length of uniformity and $a$ is the
mean distance between spins, one can limit itself by the summation
over nearest neighbours only. As an illustration, let us assume cubic
symmetry, when there are six nearest neighbours. Then the tensors,
defined in Eq. (9), are
\be
\label{38}
\{ a_{ij}\} = \frac{\mu_0^2}{a^3} \; \{ 1,1,1,1,-2,-2\} \; , \qquad
\{ b_{ij}\} = \frac{3\mu_0^2}{4a^3}\; \{ -1,-1,1,1,0,0 \} \; ,
\ee
while $c_{ij}=0$. The Fourier transforms (29) become
$$
a_k =\frac{2\mu_0^2}{a^3} \; ( \cos k_x a + \cos k_y a - 2\cos k_z a )
\; , \qquad
b_k = -\; \frac{3\mu_0^2}{2a^3}\; (\cos k_x a - \cos k_y a ) \; ,
$$
which, in the case of Eq. (37), gives
\be
\label{39}
a_k \simeq - \; \frac{\mu_0^2}{a} \; \left ( k_x^2 + k_y^2 -
2k_z^2 \right ) \; , \qquad b_k \simeq \frac{3\mu_0^2}{4a} \;
\left ( k_x^2 - k_y^2 \right ) \; .
\ee
In this manner, Eq. (36) results in the spectrum equation
\be
\label{40}
\om_k^2 \simeq \om_0^2 + \; <I_i^z>\; \frac{\mu_0^3B_0}{a\hbar^2}\;
\left ( k_x^2 + k_y^2 -2k_z^2\right ) \; .
\ee
The spectrum $\om_k$, to be positive at the maximal wave vector
$k=a^{-1}$, requires that
$$
\frac{2\mu_0^2 I}{\hbar a^3 \om_0} < 1 \; .
$$
In the standard situation, when $\mu_0^2/\hbar a^3\sim 10^5$ s$^{-1}$ and
$\om_0\sim 10^8$ Hz, this inequality holds true. The attenuation of spin
waves is of order $\Gm_2\sim\mu_0^2/\hbar a^3\sim 10^5$ s$^{-1}$, hence
it is much less than $\om_0\sim 10^8$ Hz. Thus, nuclear spin waves are well
defined.

It is necessary to stress the importance of an external magnetic field for
the stability of nuclear spin waves. Really, setting $B_0=0$ in condition
(35), we have $|a_k|>4|b_k|$. For a cubic structure with the wave vectors
in the interval (37), this yields
$$
\left | k_x^2 + k_y^2 - 2k_z^2 \right | > 3\left | k_x^2  -
k_y^2 \right | \; .
$$
This inequality should be valid for all $\bk=\{ k_x,k_y,k_z\}$
from the given interval. However, as is evident, the inequality
does not hold if either $k_x=k\equiv|\bk|$ or $k_y=k$. Hence,
these nuclear spin waves are not stable without an external
magnetic field.

One could remember that at very low temperatures, of order
$T\sim 10^{-7}$ K, nuclear spins interacting through dipolar
forces can become magnetically ordered [23]. Then in a spin system
with long-range magnetic order, there should appear spin waves, even
without external fields. This, really, may happen, but only for those
structures that are able to support nuclear spin waves. Generally,
a stable structure is characterized by two types of stability,
thermodynamic and dynamic [95]. The structure is thermodynamically
stable, when its thermodynamic potential is extremal. For instance,
when free energy is minimal. And the structure is dynamically stable
if its spectrum of collective excitations is non-negative. This is
also true for magnetic structures formed by dipolar spin systems. Only
those magnetically ordered structures will survive, which possess the
minimal free energy and are stable against arising spin waves.

As has been demonstrated above, nuclear spin waves may exist even when
there is no long-range magnetic order caused by internal forces, but
provided that there is a sufficiently strong external magnetic field.
The latter can stabilize nuclear spin waves for any kind of crystalline
symmetry. Moreover, the spin sample can be polycrystalline, being
composed of many small crystals, or even it can be amorphous. Then not
all space of the sample will be filled by a coherent system of spin
waves. But the whole sample will be separated into regions, inside
each of which spin waves are coherent, though are not coherent between
different regions. The effective linear size of such regions, where
spins can form coherent spin waves, $L_{eff}$, can be called the
uniformity length. If spin waves arise inside the spatial regions of
volume $L_{eff}^3$ which is much smaller than the total sample volume,
then such waves may be termed {\it local spin waves}. The coherence
length $L_{eff}$ must be much larger than the mean interspin distance
in order that spin waves could arise, though this length may be smaller
than the linear size of the total sample.

\section{Scale Separation Approach}

To study nonstationary spin dynamics, we have to deal with the
general equations of motion (23). In order to analyse these
equations, we invoke the scale separation approach [9,42,56,57]
consisting of three main steps: Randomization of local fields;
Classification of relative quasi-invariants; and Method of
stochastic averaging.

Let us recall that there exist the operator constructions $\xi_0$
and $\xi$, playing the role of local fields (21) or, according to
Eq. (26), of local field fluctuations. We may consider these local
fields as a separate kind of operators. Since these fields describe
local fluctuations, they can be modelled by random variables [12,53].
In this way, we have two types of variables characterizing the system,
spin operators $I_i^-,\; I_i^+,\; I_i^z$ and random fields $\xi_0,\;
\xi,\; \xi^*$. The former are responsible for long-range global
phenomena while the latter, for short-range local fluctuations.
Having defined two types of variables, we may introduce two sorts
of averaging with respect to the corresponding variables. Thus, the
statistical averaging over spin operators will be denoted by the
single angle brackets $<\ldots>$ and the averaging over random local
fields will be denoted by means of the double angle brackets
$\ll\ldots\gg$. This modelling of local fluctuating fields by random
variables is the {\it randomization of local fields}. To make the
method self-consistent, it is necessary to define the stochastic
averages over the random fields. Treating the latter as white noise,
we define
$$
\ll \xi_0(t)\gg \; = \; \ll \xi(t)\gg \; = 0 \; , \qquad
\ll \xi_0(t)\;\xi_0(t')\gg \; = 2\Gm_3\; \dlt(t-t') \; ,
$$
\be
\label{41}
\ll \xi_0(t)\;\xi(t')\gg \; = \; \ll \xi(t)\;\xi(t')\gg \; = 0\; ,
\quad \ll \xi^*(t)\;\xi(t')\gg \; = 2\Gm_3\; \dlt(t-t') \; ,
\ee
where $\Gm_3$ is the width of {\it dynamic broadening}, which is
the inhomogeneous broadening caused by the local field fluctuations.
This broadening is similar to that arising in optical resonant
systems because of dipole-dipole interactions through the exchange
of transversely polarized photons [96].

Averaging over spin operators, we could employ the decoupling
\be
\label{42}
< I_i^\al\; I_j^\bt>\; = \; <I_i^\al><I_j^\bt> \qquad (i\neq j) \; .
\ee
This looks as the mean-field approximation. However, we should
remember that the averaging, denoted by the single brackets $<\ldots>$,
by definition does not involve the stochastic variables. Therefore
the quantum correlations are not lost in the decoupling (42) but are
preserved because of the dependence of the averages $<I_i^\al>$ on
the random variables $\xi_0$ and $\xi$. This decoupling (42) may be
termed the {\it stochastic mean-field approximation} [9]. Moreover,
after having agreed to treat the spin operators and the local fields
(21)  as different variables, we do not often need the explicit usage
of the decoupling (42). This will be used only once in the definition
of coherence intensity.

Let us average the operator equations (23) over the spin operators,
without touching the local random fields (21). And let us introduce
the following definitions. The {\it transition function}
\be
\label{43}
u \equiv \frac{1}{IN} \; \sum_{i=1}^N < I_i^->
\ee
describes the average rotation of transverse spin components. The
{\it coherence intensity}
\be
\label{44}
w \equiv \frac{1}{I^2 N(N-1)} \; \sum_{i\neq j}^N <I_i^+\; I_j^- >
\ee
shows the degree of coherence in the spin motion. And the {\it spin
polarization}
\be
\label{45}
s \equiv \frac{1}{IN} \; \sum_{i=1}^N <I_i^z>
\ee
defines the average polarization in the system. Note that, under the
validity of the decoupling (42), and for $N\gg 1$, we have $w=|u|^2$,
with $u$ given in Eq. (43).

Assume that the longitudinal magnetic field $B_0$ is constant and
directed so that
\be
\label{46}
\mu_0 \; B_0 < 0 \; .
\ee
For nuclei with $\mu_0>0$, this means that $B_0<0$. Conversely, if
$\mu_0<0$, then $B_0>0$. To allow for the influence of lattice, account
must be taken of spin-lattice relaxation with a related relaxation
parameter $\Gm_1\equiv T_1^{-1}$. And to take into account the spin-spin
attenuation, we include the corresponding transverse width $\Gm_2\equiv
T_2^{-1}$.

In this way, for the variables (43) to (45), we obtain the evolution
equations
$$
\frac{du}{dt} = - i \left ( \om_0 +\xi_0 - i\Gm_2\right ) u + fs \; ,
\qquad
\frac{dw}{dt} = - 2\Gm_2 w + \left ( u^* \; f + f^*\; u\right ) s \; ,
$$
\be
\label{47}
\frac{ds}{dt} = -\; \frac{1}{2} \left ( u^*\; f + f^*\; u\right ) -
\Gm_1(s -\zeta ) \; ,
\ee
where $\om_0$ is the Zeeman frequency (34) and $\zeta\in[-1,1]$ is
an average spin polarization. In the absence of pumping, $\zeta=-1$.
But if there is a kind of pumping, for instance, by means of dynamic
nuclear polarization, then $\zeta>-1$.

Equations (47) are stochastic differential equations, containing the
stochastic variables $\xi_0$ and $\xi$. The presence of the latter
will make it possible to consider quantum effects. Also, Eqs. (47)
are nonlinear due to the resonator feedback field entering through
the effective force (22). The resonator field is defined by the
integral presentation (20), in which the magnetization density (13)
writes
\be
\label{48}
m_x = \frac{1}{2}\; \rho\; \mu_0\; I(u^* + u) \qquad
\left ( \rho \equiv \frac{N}{V} \right ) \; .
\ee
Thus, the spin evolution is characterized by the system of stochastic
nonlinear integro-differential equations (20), (22), (47), and (48).
These equations are assumed to be complimented by the initial
conditions $u(0)=u_0$, $w(0)=w_0$, and $s(0)=s_0$.

To proceed further, we notice that there exist several small
parameters. Thus, the spin-lattice and spin-spin attenuations are
usually small as compared to the Zeeman frequency, similarly to the
dynamic broadening,
\be
\label{49}
\frac{\Gm_1}{\om_0} \ll 1 \; , \qquad \frac{\Gm_2}{\om_0} \ll 1 \; ,
\qquad \frac{\Gm_3}{\om_0} \ll 1 \; .
\ee
The last inequality means that the influence of local random fields
can be treated as weak, since the stochastic averages (41) are
proportional to $\Gm_3$. The interaction energy of magnetic moments
$\mu_n\equiv\hbar\gm_n I$ with the resonator field is proportional
to the {\it natural width}
\be
\label{50}
\Gm_0 \equiv \frac{\pi}{\hbar} \; \eta \rho \mu_0^2 I =
\pi\eta\rho\gm_n\mu_n \; ,
\ee
which, being of order $\Gm_2$, is also small with respect to $\om_0$.
And the resonant circuit is assumed to be of good quality, i.e.
$Q\gg 1$, because of which the resonator ringing width $\gm$ is much
smaller than the circuit natural frequency $\om$. Hence, there are
two other small parameters
\be
\label{51}
\frac{\Gm_0}{\om_0} \ll 1 \; , \qquad \frac{\gm}{\om} \ll 1 \; .
\ee

Specifying the external transverse magnetic field, entering the total
field (3), we take
\be
\label{52}
B_1= h_0 + h_1\cos\om t \; ,
\ee
where $h_0=const$. And let the resonant part of the reduced
electromotive force (17) be
\be
\label{53}
h =h_2\cos\om t \; .
\ee
Defining the quantities
\be
\label{54}
\nu_0 \equiv \frac{\mu_0 h_0}{\hbar} \; , \qquad \nu_1 \equiv
\frac{\mu_0 h_1}{2\hbar} \; , \qquad \nu_2 \equiv
\frac{\mu_0 h_2}{2\hbar} \; ,
\ee
we keep on mind that the amplitudes of all transverse fields (52) as
well as (53) are small, in the sense that
\be
\label{55}
\frac{|\nu_0|}{\om_0} \ll 1 \; , \qquad \frac{|\nu_1|}{\om_0}
\ll 1 \; , \qquad \frac{|\nu_2|}{\om_0} \ll 1 \; .
\ee

The existence of the small parameters (49), (51), and (55) allows
us to realize the {\it classification of relative quasi-invariants}.
From Eqs. (47) we infer that the transition function (43) has to be
considered as fastly varying in time, as compared to the slow
functions (44) and (45). This implies that $w$ and $s$ are {\it
temporal quasi-invariants} with respect to $u$. Fast variation means
that with a frequency of order $\om_0$ or $\om$. These two frequencies
are supposed to be close to each other, satisfying the {\it resonance
condition}
\be
\label{56}
\frac{\Dlt}{\om} \ll 1 \; , \qquad \Dlt \equiv \om -\om_0 \; ,
\ee
which can be always done by tuning the natural resonator frequency
to the Zeeman frequency.

The occurrence of the small parameters, listed above, allows us to
solve the integral resonator equation (20) by an iteration procedure,
starting with the solution of the first of Eqs. (47) taken in zero
order with respect to small parameters, that is, substituting in the
right-hand integral of Eq. (20) the form $u\simeq u_0\exp(-i\om_0 t)$.
Accomplishing the integration gives the first-order approximant
\be
\label{57}
\frac{\mu_0 H}{\hbar} =  i(\al u -\al^* u^* ) + 2\bt\cos\om t \; ,
\ee
in which $\al=\al(t)$ is the coupling function of spins with the
resonator feedback field,
\be
\label{58}
\al = \frac{\Gm_0\om_0}{\gm + i\Dlt} \left (
1 - e^{-i\Dlt t -\gm t} \right ) \; ,
\ee
and $\bt=\bt(t)$ is the coupling function of spins with the
electromotive force,
\be
\label{59}
\bt = \frac{\nu_2}{2} \left ( 1 - e^{-\gm t} \right ) \; .
\ee
The resonance condition (56) is used in deriving Eqs. (57) with (58).

An important quantity that appears is the {\it spin-resonator coupling}
\be
\label{60}
g \equiv \frac{\gm\Gm_0\om_0}{\Gm_2(\gm^2+\Dlt^2)} \; .
\ee
This parameter enters the real and imaginary parts of the coupling
function (58), so that
$$
{\rm Re}\;\al = g\Gm_2\left [ 1 - \left ( \cos\Dlt t -\;
\frac{\Dlt}{\gm}\; \sin\Dlt t\right ) e^{-\gm t} \right ] \; ,
$$
$$
{\rm Im}\;\al = -g\Gm_2\left [ \frac{\Dlt}{\gm}  - \left ( \sin\Dlt t +
\frac{\Dlt}{\gm}\; \cos\Dlt t\right ) e^{-\gm t} \right ] \; .
$$
These formulas can be simplified if the resonance is sharp, such that
\be
\label{61}
\frac{|\Dlt|}{\gm} = 2\; \frac{|\Dlt|}{\om}\; Q \ll 1 \; .
\ee
Then the coupling function (58) becomes
\be
\label{62}
\al = g\Gm_2 \left ( 1 - e^{-\gm t}\right ) \; .
\ee
This simplification is not principal but just helps us to avoid
cumbersome expressions.

Taking into account in the effective force (22) the expression (52) for the
transverse magnetic field and Eq. (57) for the resonator field, we define
\be
\label{63}
f_1 \equiv - i\nu_0 - 2i (\nu_1+\bt) \cos\om t +\xi \; .
\ee
Then Eqs. (47) can be transformed to the system of equations
$$
\frac{du}{dt} = - i (\om_0 +\xi_0) u - (\Gm_2 -\al s) u + f_1 s -
\al su^* \; ,
$$
$$
\frac{dw}{dt} = - 2(\Gm_2 -\al s) w + (u^* f_1 + f_1^* u) s -
2\al s {\rm Re}\; u^2 \; ,
$$
\be
\label{64}
\frac{ds}{dt} = -\al w - \; \frac{1}{2}\left ( u^* f_1 + f_1^* u
\right ) - \Gm_1 ( s -\zeta) + \al {\rm Re}\; u^2 \; .
\ee
Recall that, according to the classification of relative
quasi-invariants, the function $u$ is considered as fast, being
compared  with the functions $w$ and $s$. The latter are temporal
quasi-invariants with respect to $u$. The time derivatives of the
coupling functions (59) and (62) are proportional to $\gm$, because
of which $\al$ and $\bt$ are also quasi-invariants with respect to
$u$.

Following further the scale separation approach [9,42,56,57],
we may solve the system of equations (64) by the {\it method of
stochastic averaging}, which is a generalization of the averaging
technique [43] to stochastic and partial differential equations.
First, we solve the equation for the fast variable $u$ from Eqs.
(64), with all quasi-invariants kept fixed. This yields
$$
u = u_0\exp\left\{ - ( i\om_0 + \Gm_2 -\al s ) t -
i \int_0^t \xi_0 (t')\; dt' \right \} +
$$
\be
\label{65}
+ s \int_0^t f_1(t)\exp \left\{ - (i\om_0 +\Gm_2 - \al s)
(t-t') - i\int_{t'}^t \xi_0(t'')\; dt'' \right \} \; dt' \; .
\ee
Then, we substitute the solution (65) into the equations for the slow
functions $w$ and $s$, which are the second and third of Eqs. (64). After
this, we average the resulting equations over time in the infinite interval,
keeping the quasi-invariants fixed, and over the stochastic fields, employing
the averages (41). To slightly simplify the resulting equations, we take
the initial condition for the transition function $u$ in the real form
\be
\label{66}
u_0 = \frac{1}{I} \; < I_i^x(0) > \; .
\ee
Note that the counter-rotating term, containing $u^*$, in the first of Eqs.
(64) gives a small correction to the solution (65), of order $\Gm_2/\om_0$.
Hence, this negligible correction is omitted. In accomplishing the averaging
of the second and third of Eqs. (64), the nonresonant terms, proportional
to ${\rm Re}\;u^2$, also give small contributions to the right-hand sides of
these averaged equations. Taking all this into account, we could omit in
Eqs. (64) the last terms and also could take the force (63) in the form
$$
f_1 = - i\nu_0 - i(\nu_1 +\bt) e^{-i\om t} + \xi \; .
$$
This is a kind of the rotating-wave approximation or, in other words, the
resonance approximation.

Recall that, when averaging the right-hand sides of Eqs. (64), we keep all
quasi-invariants fixed. The function $\exp(-\Gm t)$ is also considered as
a quasi-invariant. The stochastic average of Eq. (65) is
\be
\label{67}
<<u>> \; = -\; \frac{\nu_0 s}{\om_0 - i\Gm} +
\frac{(\nu_1+\bt)s}{\Dlt+i\Gm}\; e^{-i\om t} + \left [
u_0 + \frac{\nu_0 s}{\om_0 - i\Gm} \; - \;
\frac{(\nu_1+\bt)s}{\Dlt+i\Gm} \right ] e^{-(i\om_0+\Gm)t} \; ,
\ee
where the {\it collective width}
\be
\label{68}
\Gm \equiv \Gm_2 + \Gm_3 - \al\; s
\ee
is introduced. Let us also define the {\it effective attenuation}
\be
\label{69}
\tilde\Gm \equiv \Gm_3 + \frac{\nu_0^2\Gm}{\om_0^2+\Gm^2} \; -
\; \frac{\nu_0(\nu_1+\bt)\Gm}{\om_0^2+\Gm^2}\; e^{-\Gm t} +
\frac{(\nu_1+\bt)^2\Gm}{\Dlt^2+\Gm^2}\left ( 1 - e^{-\Gm t}
\right ) \; .
\ee

Following the described procedure of averaging the second and third
of Eqs. (64), we come to the equations for the {\it guiding centers},
\be
\label{70}
\frac{dw}{dt} = - 2(\Gm_2 -\al s) w + 2\tilde\Gm\; s^2 \; , \qquad
\frac{ds}{dt} = -\al w - \tilde\Gm s - \Gm_1(s - \zeta) \; .
\ee
These are the main evolution equations describing the nonlinear dynamics
of nuclear spin motion. The exponential factors entering the effective
attenuation (69) characterize retardation effects that are always present
in real processes and, hence, are to be taken into account. This retardation
may essentially influence the spin dynamics.

\section{Incoherent Quantum Stage}

One of the most interesting questions in the theory of spin superradiance is:
What initiates the motion of spins when no transition coherence is imposed
on the system at $t=0$ and no external fields push spins in the transverse
direction? That is, what is the origin of pure spin superradiance? In other
words, how the coherence in spin motion develops from initially uncorrelated
spin fluctuations?

Let us note that here we are interested in the transition coherence that is
related to the arising coherent radiation. In general, one may distinguish
two types of coherence, state coherence and transition coherence [60]. A spin
ensemble possesses {\it state coherence}, when it is prepared in a spin
polarized state, with a nonzero spin polarization (45). And {\it transition
coherence} develops when the transition function (43) is nonzero, or the
coherence intensity (44) becomes noticeable. These functions (43) to (45) may
be considered as {\it dynamical order parameters} [97] characterizing the
level of state coherence or transition coherence. If the spin sample is
initially polarized, it possesses state coherence. But if no external fields
are thrust upon the system, there is no transition coherence. An intriguing
question is: How the transition coherence develops from the state coherence
in a self-organized way?

It is well known how transition coherence appears in a system of inverted
atoms [4,5]. The relaxation begins with atomic spontaneous radiation, which
is a quantum incoherent process. After the appearance of the seed radiation
field, atomic correlations start arising through the interatomic photon
exchange. Transition coherence develops as soon as the interatomic
correlations become sufficiently intensive. Then the quantum stage of
spontaneous emission changes for the coherent stage, resulting in atomic
superradiance. Since moving spins produce magnetodipole radiation, one may
ask if the latter can be the cause of collective spin motion, by analogy
with what happens in atomic systems.

\subsection{Radiation of Magnetic Dipoles}

The vector potential created by radiating objects is
\be
\label{71}
\bA_{rad}(\br,t) = \frac{1}{c} \int \bj\left ( \br',t-\;
\frac{1}{c}\; |\br-\br'|\right ) \frac{d\br'}{|\br-\br'|} \; ,
\ee
where the density of current, formed by the magnetization density,
writes
\be
\label{72}
\bj(\br,t) \equiv c\vec\nabla\times{\bf m}(\br,t) \; .
\ee
Let this current be produced by $N_{eff}$ spins in the volume
$L_{eff}^3$. These spins can act as an altogether only if the
radiation wavelength is much larger than $L_{eff}$, so that
\be
\label{73}
k_0 L_{eff} \ll 1 \qquad \left ( k_0 \equiv \frac{\om_0}{c}
\right ) \; .
\ee
The self-action of a radiating spin corresponds to $N_{eff}=1$,
with $L_{eff}$ being the nucleus radius.

Under condition (73), the vector potential (71) can be presented
as an expansion
\be
\label{74}
\bA_{rad}(\br,t) \simeq \frac{1}{c} \int \left (
\frac{1}{x} \; - \; \frac{1}{c}\; \frac{\prt}{\prt t} +
\frac{x}{2c^2} \frac{\prt^2}{\prt t^2}\; - \; \frac{x^2}{6c^3}\;
\frac{\prt^3}{\prt t^3} \right ) \bj (\br +{\bf x},t)\; d{\bf x}
\ee
in which ${\bf x}\equiv\br'-\br$ and $x\equiv|{\bf x}|$. Assuming
that there is no current through the surface of the volume $L_{eff}^3$,
for the magnetic field ${\bf H}_{rad}\equiv\vec\nabla\times\bA_{rad}$,
we have
\be
\label{75}
{\bf H}_{rad}(\br,t) = \frac{1}{c} \int \left (
\frac{1}{x^3} \; - \; \frac{1}{2xc^2}\; \frac{\prt^2}{\prt t^2} +
\frac{1}{2c^3}\; \frac{\prt^3}{\prt t^3} \right ) \left [
{\bf x}\times \bj (\br +{\bf x},t) \right ]\; d{\bf x} \; .
\ee
The latter, involving Eq. (72) and defining the total magnetization
\be
\label{76}
{\bf M} \equiv \frac{1}{2c} \int \br \times \bj\; d\br =
\int{\bf m}\; d\br \; ,
\ee
can be transformed to
\be
\label{77}
{\bf H}_{rad}(\br,t) = \int \left [
\frac{3({\bf m}\cdot {\bf x}){\bf x}}{x^5}\; - \; \frac{{\bf m}}{x^3}
\right ] \; d{\bf x} - \; \frac{1}{2c^2} \int \left [
\frac{(\ddot{\bf m}\cdot{\bf x}){\bf x}}{x^3} +
\frac{\ddot{\bf m}}{x} \right ] \; d{\bf x} + \frac{2}{3c^3}\;
\frac{d^3{\bf M}}{dt^3} \; ,
\ee
where the overdot implies the time differentiation and ${\bf m}=
{\bf m}(\br+{\bf x},t)$.

Note that this way of presenting the magnetic field created by a system
of radiating magnetic dipoles goes back to Ginzburg [98] and since then
has been considered by many authors (see e.g. [99,100]). The first term
in Eq. (77) describes a dipolar demagnetizing field, depending on the
shape of the sample. For the spherical shape, or for a sufficiently
large volume, this term is zero, similarly to the first of Eqs. (6).
The second integral in Eq. (77) can be approximately presented [100]
as
$$
\int \left [ \frac{(\ddot{\bf m}\cdot{\bf x}){\bf x}}{x^3} +
\frac{\ddot{\bf m}}{x}\right ] \; d{\bf x} \cong
\frac{8}{3\pi L_{eff}} \; \frac{d^2{\bf M}}{dt^2} \; .
$$
Then one has
\be
\label{78}
{\bf H}_{rad} = -\; \frac{4}{3\pi c^2L_{eff}}\; \ddot{\bf M} +
\frac{2}{3c^3}\; \frac{d^3{\bf M}}{dt^3} \; .
\ee
Here the term with three time derivatives corresponds to the so-called
electromagnetic friction. Generally, such terms with odd number of
time derivatives lead to some kind of relaxation in the spin motion.
For example, the term $\dot{\bf M}$ in the Landau-Lifshitz equation
[101] can be connected [102] with spin relaxation due to spin-phonon
interactions.

The radiation field (78), in which
\be
\label{79}
{\bf M} = \mu_0 \sum_{i=1}^{N_{eff}} < {\bf I}_i> \; ,
\ee
has to be added to the total magnetic field (3). Then in the Zeeman
term (2) there appears the additional interaction $-\mu_0{\bf H}_{rad}
\cdot{\bf I}_i$. Wishing to concentrate the consideration on the role of
the radiation field (78), we assume that there are no transverse fields,
that is, $B_1=0$ and $H=0$. Then the Zeeman interaction (7) can be
written as
\be
\label{80}
\hat H_i = -\mu_0\left ( B_0 + H_{rad}^z\right ) I_i^z - \;
\frac{1}{2}\; \mu_0 \left ( H_{rad}^+\; I_i^- +
I_i^+\; H_{rad}^- \right ) \; ,
\ee
where $H_{rad}^\pm\equiv H_{rad}^x\pm i H_{rad}^y$ is expressed through
Eq. (78) and the magnetization components
$$
M^- = \mu_n N_{eff}u \; , \qquad M^z = \mu_n N_{eff} s\; ,
$$
in which $M^+$ is the complex conjugate of $M^-$ and $\mu_n\equiv\mu_0 I=
\hbar\gm_n I$. Introducing the notation for the {\it radiation width}
\be
\label{81}
\Gm_{rad} \equiv \frac{2}{3}\; k_0^3 \gm_n\mu_n N_{eff} \; ,
\ee
we may write
$$
\frac{1}{\hbar} \; \mu_0 H_{rad}^- = \frac{\Gm_{rad}}{\om_0^3}\left (
\frac{d^3 u}{dt^3} -\; \frac{2c}{\pi L_{eff}}\; \ddot{u}\right ) \; ,
\qquad
\frac{1}{\hbar} \; \mu_0 H_{rad}^z = \frac{\Gm_{rad}}{\om_0^3}\left (
\frac{d^3 s}{dt^3} -\; \frac{2c}{\pi L_{eff}}\; \ddot{s}\right ) \; .
$$
Now, the evolution equations for the functions (43) to (45) can again
be presented in the form (47), in which $\om_0$ is to be replaced by
$\om_0+\mu_0H_{rad}^z/\hbar$ and
$$
f = -\; \frac{i}{\hbar}\; \mu_0 H_{rad}^- \; .
$$
Keeping in mind the existence of the small parameter
\be
\label{82}
\frac{\Gm_{rad}}{\om_0} \ll 1 \; ,
\ee
we can find the radiation field (78) by iterating its right-hand side
with the zero-order approximation $u\simeq u_0\exp(-i\om_0 t)$ and
$s\simeq s_0$. Then
$$
f =\Gm_{rad}\left ( 1 - \; \frac{2i}{\pi k_0 L_{eff}}\right ) u
$$
and $H_{rad}^z = 0$. As a result, instead of Eqs. (47), we obtain
$$
\frac{du}{dt} = - i (\om_0 +\dlt\om_0 + \xi_0 - i\Gm_2 ) u +
\Gm_{rad} su \; ,
$$
\be
\label{83}
\frac{dw}{dt} = - 2(\Gm_2 - \Gm_{rad} s) w \; , \qquad
\frac{ds}{dt} = - \Gm_{rad} w  -\Gm_1 (s-\zeta) \; ,
\ee
with the frequency shift
$$
\dlt \om_0 \equiv \frac{2\Gm_{rad} s}{\pi k_0 L_{eff}} \; .
$$

Equations (83) show that the magnetodipole radiation can lead to the
arising collective effects only if the radiation width (81) is much
larger than the transverse dephasing width $\Gm_2$ due to dipole-dipole
interactions. For the latter one has
\be
\label{84}
\Gm_2 = n_0 \rho\gm_n \mu_n \; ,
\ee
where $n_0$ is of the order of the nearest-neighbour number [12,23,53].
Comparing Eqs. (81) and (84), with taking account of
$\rho=N_{eff}/L_{eff}^3$, we find
\be
\label{85}
\frac{\Gm_{rad}}{\Gm_2} = \frac{2}{3n_0}\left ( k_0 L_{eff}
\right )^3 \ll 1 \; ,
\ee
since $n_0\sim10$ and $k_0L_{eff}\ll 1$ according to condition (73).
Hence $\Gm_{rad}\ll\Gm_2$ and, respectively, the radiation time
$T_{rad}\equiv\Gm_{rad}^{-1}$ is much larger then the dephasing time
$T_2\equiv\Gm_2^{-1}$.

To estimate the related parameters, we may take the typical value
$\om_0\sim 10^8$ Hz, so that $k_0\sim 10^{-2}$ cm$^{-1}$. For protons,
the gyromagnetic ratio $\gm_n=2.675\times 10^4$ G$^{-1}$s$^{-1}$ and
spin $I=1/2$. The proton magnetic moment $\mu_n=2.793\mu_N$ can be
expressed through the nuclear magneton $\mu_N=5.051\times 10^{-24}$
erg G$^{-1}$. We have $\mu_n=1.411\times 10^{-23}$ erg G$^{-1}$. The
dimension of Gauss is such that G$^2$=erg$\cdot$cm$^{-3}$. In this
way, $\Gm_{rad}\sim 10^{-25}N_{eff}$ s$^{-1}$, while $\Gm_2\sim 10^5$
s$^{-1}$ for $\rho\sim 10^{23}$ cm$^{-3}$. The relaxation caused by
the self-action through magnetodipole radiation, when $N_{eff}=1$,
yields the radiation time $T_{rad}\sim 10^{25}$ s. This, with 1 year
$\sim 10^7$ s, gives $T_{rad}\sim 10^{18}$ years. Such an enormous
time is not only much larger than $T_2\sim 10^{-5}$ s, but also
surpasses the Earth lifetime of $5\times 10^9$ years and is even
longer than the Universe lifetime of 10$^{10}$ years. Clearly, the
self-action, caused by the magnetodipole radiation, is completely
negligible. Even with the account of collective effects, when
$N_{eff}\sim 10^{23}$, we get $\Gm_{rad}\sim 10^{-2}$ s$^{-1}$,
hence $T_{rad}\sim 10^2$ s, which is much larger then $T_2$, their
ratio being $T_{rad}/T_2\sim 10^7$. These estimates show that the
ratio (85) is extremely small, $\Gm_{rad}/\Gm_2\leq 10^{-7}$.

Thus, we come to the conclusion that magnetodipole radiation is
absolutely unable to organize the coherent motion of spins. This
conclusion is based on the inequality (85), which is always valid,
independently of the nature of the spins involved, whether these are
nuclear, electron, atomic, or molecular spins.

\subsection{Resonator Nyquist Noise}

For many years, practically all researches have used to write that
the major cause producing spin motion, resulting in pure superradiance
is the thermal Nyquist noise of the resonant electric circuit. This
belief has been especially surprising because Bloembergen and Pound
[7] mentioned that this noise cannot be a noticeable cause of spin
relaxation. The common delusion about the principal role of the Nyquist
noise was clarified in the papers [10,55--57].

The role of this noise can be studied by analysing the effective
attenuation (69). Assume that there are no transverse external
fields, so that $\nu_0=\nu_1=0$, but there is only the
electromotive force corresponding to the resonator Nyquist noise.
Then the effective attenuation (69) is
\be
\label{86}
\tilde\Gm = \Gm_3 + \Gm_{res} \; ,
\ee
where, in accordance with Eq. (59),
\be
\label{87}
\Gm_{res} = \frac{\nu_2^2\Gm}{4(\Dlt^2+\Gm^2)} \left ( 1 -
e^{-\Gm t}\right ) \left ( 1 - e^{-\gm t}\right )^2 \; .
\ee
The latter attenuation is caused by the Nyquist noise of the resonant
electric circuit. For concreteness, we shall keep in mind the standard
situation, when the resonator width $\gm$ is larger or of the order of
$\Gm_2$ and $\Gm_3$. And, for simplicity, we consider the case of good
resonance, when condition (61) is valid. Then at short time, the
attenuation (87) reads
\be
\label{88}
\Gm_{res} \simeq \frac{1}{4}\; \nu_2^2 \gm_2 t^3 \qquad
(\gm t\ll 1) \; .
\ee
Notice that $\Gm_{res}=0$ at $t=0$.

The resonator electromotive force, entering the Kirchhoff equation
(11), can be written as
\be
\label{89}
E_f = E_{res}\; \cos\om t \; .
\ee
The amplitude of the reduced electromotive force (17) or (53) is $h_2$,
for which we have
$$
h_2 = \frac{cE_{res}}{nA_c\gm} \; , \qquad h_2^2 =
\frac{8\pi\eta E_{res}^2}{\gm RV} \; .
$$
The amplitude squared of the electromotive force (89) caused by the
Nyquist noise, can be presented [18] as
\be
\label{90}
E_{res}^2 = \frac{\hbar\om}{2\pi} \; \gm R \; {\rm coth}\;
\frac{\om}{2\om_T} \qquad \left ( \om_T \equiv \frac{k_BT}{\hbar}
\right ) \; ,
\ee
with $\om_T$ being the thermal frequency. At radiofrequencies, one has
$\om\ll\om_T$. Then Eq. (90) can be slightly simplified to
\be
\label{91}
E_{res}^2 \simeq \frac{\hbar}{\pi} \; \gm R \; \om_T \qquad
\left ( \frac{\om}{\om_T} \ll 1 \right ) \; .
\ee
For the quantity $\nu_2$, defined in Eq. (54), we get
\be
\label{92}
\nu_2^2 = \frac{2\Gm_0E_{res}^2}{\hbar\gm IRN} \simeq
\frac{2\Gm_0\om_T}{\pi IN} \; .
\ee
For the attenuation (88) at $t\approx 1/\gm$, we find
\be
\label{93}
\Gm_{res} \simeq \frac{\nu_2^2}{4\gm} \simeq
\frac{\Gm_0\om_T}{2\pi I\gm N} \; .
\ee
This is, actually, the maximal value of the relaxation that can be
reached by the attenuation (87).

To estimate Eq. (93), we again accept the typical values of
$\Gm_2\sim\Gm_0\sim 10^5$ s$^{-1}$ and $\gm\sim 10^6$ s$^{-1}$.
At temperature $T\sim 0.1$ K, one has $\om_T\sim 10^{11}$ Hz. Thus,
we obtain $\Gm_{res}\sim (10^{10}/N)$ s$^{-1}$. If $N\sim 10^{23}$,
then $\Gm_{res}\sim 10^{-13}$ s$^{-1}$, which is much smaller than
$\Gm_3\sim\Gm_2$. The related time, during which the Nyquist noise
could produce spin relaxation, $T_{res}\equiv\Gm_{res}^{-1}$, is
$T_{res}\sim 10^{13}$ s $\sim 10^6$ years. This is many orders
longer that $T_2\sim T_3\equiv\Gm_3^{-1}$. Therefore, the resonator
Nyquist noise plays no role in spin relaxation of macroscopic samples
and can never serve as a triggering cause starting the spin motion,
unless the number of spins is much smaller than $10^5$.

\subsection{Local Spin Fluctuations}

We continue considering the case when there are no transverse external
fields pushing spins, so that $\nu_0=\nu_1=0$. As we have found out,
neither the magnetodipole radiation, nor the resonator Nyquist noise
are able to start the spin motion. So, we set $\nu_2=0$, hence $\bt=0$.
Then the effective attenuation (69) becomes $\tilde\Gm=\Gm_3$. Recall
that the width $\Gm_3$ is due to the dynamic broadening caused by
local spin fluctuations, which are a kind of local spin waves. These
spin fluctuations are, thus, the sole possible origin that could
trigger the spin motion, when there are no external fields.

To understand how the spin motion starts, consider short times, when
$\gm t\ll 1$ and the resonator coupling function (62) is yet small,
$\al\approx 0$. Then Eqs. (70), under $\tilde\Gm=\Gm_3$, are
\be
\label{94}
\frac{dw}{dt} =-2\Gm_2 w + 2\Gm_3 s^2 \; , \qquad
\frac{ds}{dt} = -(\Gm_1+\Gm_3) s +\Gm_1 \zeta \; .
\ee
Their solution at short times reads
$$
w\simeq \left ( w_0 -\; \frac{\Gm_3}{\Gm_2}\; s_0^2\right ) \exp
( - 2\Gm_2 t) + \frac{\Gm_3}{\Gm_2}\; s_0^2 \; ,
$$
\be
\label{95}
s \simeq \left ( s_0 -\; \frac{\Gm_1\zeta}{\Gm_1+\Gm_3} \right )
\exp\{ -(\Gm_1+\Gm_3)t\} + \frac{\Gm_1\zeta}{\Gm_1+\Gm_3} \; .
\ee
This shows that, even if no transverse coherence is imposed on
the spin system at the initial time, so that $w_0=0$, the coherence,
anyway, begins to arise triggered by the local spin fluctuations in
the presence of a nonzero initial polarization.

The resonator coupling function (62) increases with time. The incoherent
quantum stage of spin motion lasts till that time when the coupling
function $\al(t)$ reaches a value such that $\al s=\Gm_2$. Then, as is
seen from Eqs. (70), fast generation of transverse coherence starts in
the system. The crossover time $t_c$, separating the incoherent quantum
stage and the coherent regime of motion, is defined by the equality
\be
\label{96}
\al(t_c)s(t_c)= \Gm_2 \; .
\ee
Assuming that the crossover time $t_c$ is the smallest among other
characteristic relaxation times, such as $\tau$, $T_1$, and $T_2$, we
can simplify the solution (95) to the form
\be
\label{97}
w(t_c) \simeq w_0 + 2\Gm_3 t_c s_0^2 \; , \qquad
s(t_c) \simeq s_0 + 2\Gm_1 t_c \zeta \; ,
\ee
taken at $t=t_c$. This assumption about the time $t_c$ being the
shortest is necessary for the existence of superradiance. From
definition (96), we find the {\it crossover time}
\be
\label{98}
t_c = \tau\ln\left ( \frac{gs_0}{gs_0-1} \right ) \; ,
\ee
where $\tau$ is the resonator ringing time. In order that the quantum
stage could certainly change for the coherent regime, the crossover
time (98) must be positive and finite, which requires that
\be
\label{99}
g s_0 > 1 \qquad (t_c>0) \; ,
\ee
which is, actually, the condition of maser generation. In the case
of sufficiently strong coupling, Eq. (98) reduces to
\be
\label{100}
t_c \simeq  \frac{\tau}{g s_0} \qquad (g s_0 \gg 1 ) \; ,
\ee
which shows that the crossover time can really be made shorter than
other characteristic times. Since the spin-resonator coupling (60)
is positive, condition (99) also tells us that the initial spin
polarization $s_0$ is to be positive, which implies that the spin
system is to be prepared in a nonequilibrium state, being in an
external field satisfying condition (46).

\section{Regimes of Coherent Radiation}

After the crossover time (98), the coupling function (62) quickly
increases and reaches the value $\al\approx g\Gm_2$. At this stage,
the motion of spins becomes coherent. The main regimes of coherent
spin radiation are described below.

\subsection{Transient Spin Superradiance}

The transient stage corresponds to times larger than the crossover
time $t_c$ but essentially shorter than the longitudinal relaxation
time $T_1$. Then in Eqs. (70) one may neglect the longitudinal
relaxation parameter $\Gm_1$. Consider the case, when there are
no external transverse fields, so that the effective attenuation
(69) is $\tilde\Gm=\Gm_3$. Here we take into account that the
resonator noise plays no role in spin motion. Being interested in
the situation of sufficiently strong spin-resonator coupling $g\gg 1$,
we see that $g\Gm_2\gg\Gm_3$, since $\Gm_2\sim\Gm_3$. Thence, we
may omit in Eqs. (70) the terms containing $\Gm_3$. Thus, at this
transient stage, we consider the equations
\be
\label{101}
\frac{dw}{dt} = - 2\Gm_2( 1 - g s) w \; , \qquad
\frac{ds}{dt} = - g\Gm_2 w \; .
\ee
These equations possess an exact solution [10,55--57] that reads
\be
\label{102}
w=\left (\frac{\Gm_p}{g\Gm_2}\right )^2 \; {\rm sech}^2
\left ( \frac{t-t_0}{\tau_p}\right ) \; , \qquad
s=-\; \frac{\Gm_p}{g\Gm_2}\; {\rm tanh} \left (
\frac{t-t_0}{\tau_p}\right ) + \frac{1}{g} \; .
\ee
The integration parameters in this solution are obtained by considering
as the initial condition the values (97) at $t=t_c$, which gives the
{\it delay time}
\be
\label{103}
t_0 = t_c +\frac{\tau_p}{2}\; \ln \left |
\frac{\Gm_p+\Gm_g}{\Gm_p-\Gm_g}\right | \; ,
\ee
where
\be
\label{104}
\Gm_p^2 = \Gm_g^2 +(g\Gm_2)^2\left ( w_0 + 2\Gm_3 t_c s_0^2\right ) \; ,
\qquad \Gm_g \equiv \Gm_2(gs_0-1)\; , \qquad \Gm_p\tau_p = 1\; .
\ee
In order that the delay time be larger than the crossover time, but finite,
that is, $t_c<t_0<\infty$, it should be that $\Gm_p>\Gm_g>0$, which yields
\be
\label{105}
w_0 + 2\Gm_3 t_c s_0^2 > 0 \; , \qquad gs_0 > 1 \; .
\ee
At the delay time (103), the coherence intensity is maximal,
\be
\label{106}
w(t_0)  =\left ( s_0 -\; \frac{1}{g}\right )^2
( 1 + 2\Gm_3 t_c)\; , \qquad s(t_0) = \frac{1}{g} \; ,
\ee
after which the transition coherence fastly decays,
\be
\label{107}
w\simeq 4 w(t_0)\exp(-2\Gm_p t) \; , \qquad
s\simeq -s_0 + \frac{2}{g} \qquad (t\gg t_0) \; .
\ee

The effective time of the coherent pulse, keeping in mind that $\Gm_3
t_c\ll 1$, writes
\be
\label{108}
\tau_p = \frac{T_2}{\sqrt{(gs_0-1)^2+g^2w_0}} \left [ 1 - \;
\frac{\Gm_3 t_c(gs_0)^2}{(gs_0-1)^2+g^2w_0}\right ] \; .
\ee
The necessary condition for superradiance is $\tau_p<T_2$, from which
one has the inequality
\be
\label{109}
(gs_0 -1 )^2 + g^2 w_0 > 1 \; .
\ee
However, condition (109) is not sufficient for superradiance and defines
three possible regimes:
$$
w_0 = 0 \; , \qquad gs_0 > 2 \qquad (pure \; superradiance) \; ,
$$
$$
g^2w_0> 1- (gs_0 - 1)^2 \; , \qquad gs_0 > 1 \qquad (triggered \;
superradiance) \; ,
$$
\be
\label{110}
g^2 w_0 > 1 \; , \qquad gs_0 \leq 1 \qquad (collective\;
induction) \; .
\ee
Recall that superradiance is a self-organized process, which requires
that the crossover time $t_c$ be positive, hence, according to Eq. (99),
it should be that $gs_0>1$. Collective induction is not a self-organized
process, but it is induced by an initially imposed coherence intensity
$w_0$.

Thus, there exist two types of transient spin superradiance, pure
superradiance that is developing as a completely self-organized process
and triggered superradiance when the process is mainly self-organized but
at the beginning it is slightly pushed by an externally imposed coherence.
As follows from Eqs. (107), it is only in the regime when $gs_0>2$ that
the spin polarization becomes inverted after the superradiant pulse. This
may happen in either the regime on pure or triggered superradiance. When
$g\gg 2$, the polarization $s$ after $t_0$ is almost completely inverted
to the value $-s_0$. This effect of polarization reversal can be used for
fast repolarization of solid-state targets employed in scattering experiments
[29,103].

\subsection{Pulsing Spin Superradiance}

If there are no external fields acting on the spin system, then, after
the transient superradiant burst, occurring at the delay time $t_0$,
the transition coherence dies out. In order to produce a series of
superradiant pulses, it is necessary to involve an external action.
There are two ways of organizing a regime with a series of superradiant
bursts, which can be called pulsing superradiance and punctuated
superradiance.

To realize the regime of pulsing spin superradiance, the inversion of
spin polarization has to be supported by a permanent pumping, which can
be accomplished by means of dynamic nuclear polarization. In that case,
the longitudinal relaxation parameter $\Gm_1$ plays the role of the pump
rate $\Gm_1^*$ that can be much larger than the spin-lattice attenuation
of $10^{-5}$ s$^{-1}$. The pump rate $\Gm_1^*$ can be 10 s$^{-1}$ or
larger. In what follows, we shall continue writing $\Gm_1$ instead of
$\Gm_1^*$, keeping in mind that the value of $\Gm_1$ corresponds to the
pump rate. This is done in order to avoid cumbersome notations. In the
presence of pumping, the stationary polarization $\zeta$ is the pump
polarization, which can reach the value of $\zeta=1$.

We consider the case when there are no external transverse fields, so
that the effective attenuation (69) reduces to $\tilde\Gm=\Gm_3$. At
larger time, when $\gm t\gg 1$, the coupling function (62) acquires its
maximal value $\al\cong g\Gm_2$. Then Eqs. (70) write
\be
\label{111}
\frac{dw}{dt} = -2\Gm_2 ( 1 -gs) w + 2 \Gm_3 s^2 \; , \qquad
\frac{ds}{dt} = -g\Gm_2 w - \Gm_3 s - \Gm_1 (s-\zeta) \; .
\ee
To find the conditions when the pulsing regime could arise, we need
to consider the long-time behaviour of the solutions to Eqs. (111). For
this purpose, we define the stationary solutions to these equations and
accomplish the Lyapunov stability analysis [60]. The appearance of complex
characteristic exponents, related to particular fixed points, would mean
the existence of oscillations around the corresponding stationary
solutions.

If the pumping is not very strong, so that $g\zeta\ll -1$, then the
stable fixed points are
\be
\label{112}
w_1^* \simeq \frac{\zeta^2\Gm_3}{|g\zeta|\Gm_2} \; , \qquad
s_1^*\simeq \zeta \left ( 1 -\; \frac{\Gm_3}{|g\zeta|\Gm_1}
\right ) \; .
\ee
For the associated characteristic exponents, we find
$$
J_1^+\simeq -\Gm_1 \left ( 1 -\; \frac{\Gm_3}{2|g\zeta|\Gm_2}
\right ) \; , \qquad
J_1^- \simeq -2\Gm_2 ( 1 +|g\zeta|) -\; \frac{2\Gm_2\Gm_3}{\Gm_1}\;
\left ( 1 - \; \frac{\Gm_1}{2\Gm_2} \right ) \; .
$$
These exponents, being real and negative, show that the fixed point (112)
is a stable node.

When the spin-resonator coupling is weak or the pumping is not strong, so
that $|g\zeta|\ll 1$, then the stable stationary solution is
$$
w_1^*\simeq \left ( \frac{\zeta\Gm_1}{\Gm_1+\Gm_3}\right )^2 \;
\frac{\Gm_3}{\Gm_2}\; \left [ 1 +
\frac{\Gm_1(\Gm_1-\Gm_3)}{(\Gm_1+\Gm_3)^2} \; g\zeta \right ] \; ,
$$
\be
\label{113}
s_1^* \simeq \frac{\zeta\Gm_1}{\Gm_1+\Gm_3}\left ( 1  -\;
\frac{\Gm_1\Gm_3}{\Gm_1+\Gm_3}\; g\zeta\right ) \; .
\ee
The characteristic exponents, if $\Gm_1+\Gm_3\neq 2\Gm_2$, are
$$
J_1^+ \simeq -2\Gm_2 +
\frac{2\Gm_1\Gm_2(\Gm_1-2\Gm_2-\Gm_3)}{(\Gm_1+\Gm_3)(\Gm_1-2\Gm_2+\Gm_3)}\;
g\zeta \; ,
$$
$$
J_1^- \simeq -\Gm_1 -\Gm_3 +
\frac{4\Gm_1\Gm_2\Gm_3}{(\Gm_1+\Gm_3)(\Gm_1-2\Gm_2+\Gm_3)}\; g\zeta \; ,
$$
which classifies the stationary solution (113) as a stable node. But in the
special case, when
\be
\label{114}
\Gm_1 + \Gm_3 = 2\Gm_2 \; ,
\ee
the characteristic exponents become
$$
J_1^\pm \simeq -2\Gm_2 \pm i \left ( 2\Gm_1\Gm_3 g\zeta\right )^{1/2} +
\frac{1}{2}\; \Gm_1 g\zeta \; .
$$
Then solutions (113) describe a stable focus, if $g\zeta>0$. This means that
there exists a series of pulses approximately separated from each other by
the {\it separation time}
\be
\label{115}
T_{sep} = \pi\; \sqrt{ \frac{2T_1T_3}{g\zeta}} \qquad
(0 < g\zeta\ll 1) \; ,
\ee
where $T_3\equiv\Gm_3^{-1}$ and condition (114) is valid.

For strong spin-resonator coupling and sufficient pumping, such that
$g\zeta\gg 1$, the stationary solution to Eqs. (111) is
\be
\label{116}
w_2^* \simeq \frac{\zeta\Gm_1}{g\Gm_2} \; , \qquad s_2^*\simeq \frac{1}{g} \; .
\ee
The characteristic exponents
$$
J_2^\pm \simeq -\; \frac{1}{2}\; (\Gm_1 + \Gm_3) -\;
\frac{\Gm_2\Gm_3}{g\zeta\Gm_1}
$$
show that solution (116) is a stable node.

Now let us analyse the behaviour of stationary solutions to Eqs. (111) for
arbitrary values of $g\zeta$, but for different relations between the pumping
rate $\Gm_1$ and the dynamic broadening $\Gm_3$. Thus, when $\Gm_1\ll\Gm_3$,
then
\be
\label{117}
w_1^* \simeq \frac{\zeta^2\Gm_1^2}{\Gm_2\Gm_3} \; , \qquad
s_1^* \simeq \zeta\; \frac{\Gm_1}{\Gm_3} \left [ 1 - (1+g\zeta)\;
\frac{\Gm_1}{\Gm_3}\right ] \; .
\ee
The corresponding characteristic exponents, if $\Gm_3\neq 2\Gm_2$, are
$$
J_1^+ \simeq -2\Gm_2 - \;
\frac{2(2\Gm_2+\Gm_3)\Gm_1\Gm_2}{(2\Gm_2-\Gm_3)\Gm_3}\; g\zeta\; ,
\qquad
J_1^- \simeq -\Gm_3 - \;
\frac{(4g\zeta +2\Gm_2-\Gm_3)\Gm_1}{2\Gm_2-\Gm_3}\;  ,
$$
which defines the fixed point (117) as a stable node. And for the special
case, when
\be
\label{118}
\Gm_3 = 2\Gm_2 \; ,
\ee
one gets
$$
J_1^\pm \simeq -\Gm_3 \pm i\; \sqrt{2g\zeta\Gm_1\Gm_3} \; +
\frac{1}{2} (g\zeta -1)\Gm_1 \; .
$$
In such a case, the fixed point (117) becomes a stable focus, provided
that $g\zeta>0$. This implies the existence of a series of pulses, with
the separation intervals
\be
\label{119}
T_{sep} = \pi\; \sqrt{\frac{T_1T_2}{g\zeta}} \qquad
(\Gm_1\ll\Gm_3=2\Gm_2) \; .
\ee

Finally, for a large pump rate, such that $\Gm_1\gg\Gm_3$, we have the
stationary solutions, which acquire different forms for different values
of $g\zeta$. In the case of weak coupling, when $g\zeta<1$, we get
\be
\label{120}
w_1^* \simeq \frac{\zeta^2\Gm_3}{(1-g\zeta)\Gm_2}\; , \qquad s_1^* \simeq
\zeta \left [ 1 -\; \frac{\Gm_3}{(1-g\zeta)\Gm_1} \right ] \; ,
\ee
with the characteristic exponents
$$
J_1^+ \simeq -\Gm_1 \; , \qquad J_1^- \simeq -2(1-g\zeta)\Gm_2
$$
telling that the fixed point (120) is a stable node. For the intermediate
case, when $g\zeta=1$, we find
\be
\label{121}
w_1^* \simeq \frac{\zeta^2\sqrt{\Gm_1\Gm_3}}{\Gm_2}\; \left ( 1 - \;
\frac{3}{2}\; \sqrt{\frac{\Gm_3}{\Gm_1}} \right ) \; , \qquad
s_1^* \simeq \zeta \left ( 1 -\; \sqrt{\frac{\Gm_3}{\Gm_1}} +
\frac{\Gm_3}{2\Gm_1}\right ) \; .
\ee
This point is again a stable node since
$$
J_1^+ \simeq -\Gm_1 + 2\Gm_2\; \sqrt{\frac{\Gm_3}{\Gm_1}} \; , \qquad
J_1^- \simeq - 4\Gm_2\; \sqrt{\frac{\Gm_3}{\Gm_1}} \;
$$
And for strong coupling and effective pumping, when $g\zeta>1$, we obtain
\be
\label{122}
w_2^* \simeq \frac{(g\zeta-1)\Gm_1}{g^2\Gm_2}\; \left [ 1 - \;
\frac{(g\zeta-2)\Gm_3}{(g\zeta -1)^2\Gm_1} \right ] \; , \qquad
s_2^* \simeq \frac{1}{g} \left [ 1 -\;
\frac{\Gm_3}{(g\zeta-1)\Gm_1}\right ] \; .
\ee
The related characteristic exponents
$$
J_2^\pm \simeq -\; \frac{1}{2}\; \Gm_1 \left [ 1 \pm
\sqrt{1-8\; \frac{\Gm_2}{\Gm_1}\; (g\zeta-1)} \right ]
$$
are real in the interval $1<g\zeta<1+\Gm_1/8\Gm_2$, but become complex if
\be
\label{123}
g\zeta > 1 + \frac{\Gm_1}{8\Gm_2} \; .
\ee
Under this condition, the frequency
\be
\label{124}
\om_\infty \equiv \sqrt{2\Gm_1\Gm_2(g\zeta - 1 -\; \frac{\Gm_1}{8\Gm_2})}
\ee
is real. Then we may write
$$
J_2^\pm \simeq -\; \frac{1}{2}\; \Gm_1 \pm i\om_\infty \; .
$$
This means that the fixed point (122) is a stable focus. Here again
there arises a series of pulses separated by the time
\be
\label{125}
T_{sep} \equiv \frac{2\pi}{\om_\infty} \cong \pi\;
\sqrt{\frac{2T_1T_2}{g\zeta-1}} \; ,
\ee
where, to simplify the expression, we take into account that $\Gm_1<\Gm_2$.

In this way, the regime of pulsing superradiance can appear under the
conditions resulting in the formation of a series of pulses, which at long
times, are separated from each other by the corresponding separation time
(115), or (119), or (125), depending on the accepted conditions. This
conclusion has been confirmed by direct numerical solution of Eqs. (111),
displaying the regime of pulsing superradiance [9,63,65]. Experimental
observation of this regime was done in Refs. [30--33]. Pulsing spin
superradiance can be used for organizing superradiant operation of spin
masers [60,65].

There exists another possibility of producing a series of superradiant pulses,
which was named {\it punctuated spin superradiance} [66]. This can be done
in the following way. Suppose that the transient spin superradiance has been
realized, resulting in a superradiant burst at the delay time $t_0$. Then,
according to Eqs. (107), at some time after $t_0+\tau_p$ the spin polarization
is reversed. For large coupling $g$, this reversal is practically complete.
Now assume that we again inverse the spin polarization from $-s_0$ to $s_0$,
thus obtaining an inverted nonequilibrium system. Such an inversion can be
realized in three ways: reversing the external magnetic field $B_0$, acting
on spins by a resonant $\pi$-pulse, or turning the sample $180^o$ about the
$x$-axis. After the time $t_0$, counted from the moment when the newly
nonequilibrium state is prepared, another superradiant burst will arise.
When the second burst dies out, one can again invert the spin polarization
by one of the described methods. Then one more superradiant pulse will
occur. This procedure can be repeated as many times as required for creating
a desired number of sharp superradiant pulses. Contrary to the regime
of pulsing superradiance, analysed above, when the intervals between
superradiant bursts are defined by the system parameters, the time intervals
between superradiant  pulses in punctuated superradiance can be regulated.
Is is admissible to form various groups of pulses, with different intervals
between separated groups. Hence, it is feasible to compose a code, like the
Morse alphabet, which could be employed for processing information [66].

\subsection{Induced Coherent Emission}

In the investigation of radiation regimes above, we have assumed that there
are no external transverse fields that would permanently act on spins. The
possible existence of such external fields has been reduced only to the
preparation of initial conditions for $w_0$ and $s_0$, or to supporting
the value of the pumping parameter $\zeta$. But the external transverse
field (52) has been switched off. This was done for studying the process
of self-organization, which would not be perturbed by external fields
pushing spins in the transverse direction and, thus, helping to develop
the transition coherence. The temporal arising of self-organized coherent
motion of spins, from an initially incoherent state, is the most interesting
and the least studied problem. Actually, the gradual appearance of the
transition spin coherence has never been described before Refs. [8,60,66].
This is why we, first, paid the main attention to this problem.

Switching on the external transverse field (52) of course influences spin
dynamics. The presence of this field makes the classification of possible
regimes of coherent emission less evident. If the transverse field (52)
is very weak, so that the leading part in the effective attenuation (69)
is due to the dynamic dipolar broadening $\Gm_3$, then the analysed above
regimes do not change much. But if the transverse field is sufficiently
strong, so that in the effective attenuation (69) the term $\Gm_3$ is not
the largest, but other terms are either also essential or even prevail over
$\Gm_3$, than spin dynamics may be noticeably disturbed. However, in general,
one may distinguish two principally different regimes related to either weak
or strong spin-resonator coupling. If this coupling is weak, such that
$gs_0\leq 1$, then even very strong transverse fields can induce coherent
radiation with only $\tau_p\geq T_2$. But when both the spin-resonator
coupling is strong, so that $gs_0>1$, and the external transverse field
is strong, then the pulse of induced coherent emission can become short,
with $\tau_p<T_2$. In any case, when the spin system is subject to the
action of the transverse coherent field (52), but there is no pumping
supporting the value of the pumping parameter $\zeta>-1$, then there can
arise the sole main coherent burst, which may be accompanied by a series
of small fastly attenuating oscillations. That is, the transverse field (52)
induces only transient coherence radiation. This type of coherent radiation,
essentially caused and influenced by an external transverse field, can be
called {\it induced coherent emission}. This should not be confused with
superradiance that is a self-organized process occurring without the action
of any transverse field of type (52).

As an example, let us consider how a strong constant transverse field would
influence spin radiation. When in the transverse field (52) only the constant
part $h_0$ is present, but $h_1=0$, then the effective attenuation (69) takes
the form
$$
\tilde\Gm = \Gm_3 + \frac{\nu_0^2\Gm}{\om_0^2+\Gm^2} \; ,
$$
with $\nu_0$ defined in Eq. (54) and $\Gm$ in Eqs. (68) and (62). From here,
we see, first, that the applied constant field has to be rather strong in
order to lead to a noticeable, as compared to $\Gm_3$, term. Generally, the
second term here is smaller than $\Gm_3$, if condition (55) holds true. If
so, then the following process is not much different from what has been
analysed above. Just in all formulas, we need to shift $\Gm_3$ replacing
it by
$$
\Gm_3 + \frac{\nu_0^2(\Gm_2+\Gm_3)}{\om_0^2+(\Gm_2+\Gm_3)^2}
\cong \Gm_3 + \frac{\nu_0^2}{\om_0^2}\; (\Gm_2+\Gm_3) \; .
$$
Such a shift would mainly influence the value of the delay time (103). In
particular, the natural magnetic anisotropy of spin samples is often modelled
by an external constant field. Hence, the existence of such an anisotropy would
also play its role in triggering the initial spin motion [61,62]. However, the
term $\Gm_3$ plays the major role.

In the case, when the transverse field (52) consists of only a resonant
alternating field, with the amplitude $h_1$, then the effective attenuation
(69) becomes
$$
\tilde\Gm =\Gm_3 + \frac{\nu_1^2\Gm}{\Dlt^2+\Gm^2}\left ( 1 - e^{-\Gm t}
\right ) \; ,
$$
where $\nu_1$ is given in Eq. (54), $\Gm=\Gm_2+\Gm_3-\al s$, and the coupling
function $\al$ being defined in Eq. (62). We have thoroughly analysed the
behaviour of solutions to Eqs. (70) for various system parameters, when the
resonant transverse field is present. This has been accomplished by numerically
solving Eqs. (70) for different initial conditions $w_0$ and $s_0$. Note that
the term in $\tilde\Gm$, due to the resonant field, is of the order of
$\nu_1^2/\Gm_2$, which can be comparable with or even larger than $\Gm_3$.
Thence the resonant field influences spin motion essentially stronger than
a constant transverse field, which is, actually, rather understandable.
Numerically solving Eqs. (70), we found that, varying the system parameters
and the initial conditions, it is possible to realize a variety of transient
coherent pulses. Increasing the amplitude of the resonator field makes the
duration of the coherent pulse longer. In all the cases the solutions $w(t)$
and $s(t)$, independently of their initial conditions, dies out to almost zero
on the scale not longer than $T_2$. The attenuation of the solutions slows
down when increasing $\nu_1$. The latter, under conditions (55), is limited by
the relation $\nu_1/\Gm_2\ll 10^3$. If the external resonant field is so strong
that conditions (55) are no longer valid, then, instead of Eqs. (70), one
should go back to Eqs. (47). In the latter case of a very strong resonant
field, the solutions to Eqs. (47) may display chaotic behaviour [33,104,105].

In conclusion to this section, discussing different regimes of coherent spin
radiation, it is useful to recall how the latter could be measured. An ensemble
of coherently moving nuclear spins generates magnetodipole radiation with the
total intensity
\be
\label{126}
I(t) = \frac{2}{3c^3}\; \left | \ddot{\bf M}(t)\right |^2 \; ,
\ee
in which
$$
{\bf M} = \hbar \gm_n \sum_{i=1}^N \; < {\bf I}_i(t)>
$$
is the total magnetization of $N$ nuclei. The standardly considered quantity
is the radiation intensity averaged over fast oscillations. For the intensity
(126), this gives
\be
\label{127}
\overline I(t) = \frac{2}{3c^3}\; \mu_n^2\om_0^4 N^2 w(t) \; ,
\ee
where $\mu_n\equiv\hbar\gm_n I$ and $w(t)$ is the function of coherence
intensity (44). The proportionality of the radiation intensity (127) to the
number of spins squared is characteristic of any coherent radiation, which
is not necessarily superradiance. The level of radiation intensity (127) is
rather weak. Thus, the maximal intensity, when $w(t)\approx 1$, for proton
spins, with $\mu_n\sim 10^{-23}$ erg G$^{-1}$, $\om_0\sim 10^8$ Hz, and
$N\sim 10^{23}$, is only $\overline I\sim 10^{-5}$ W.

However, despite such weak radiation intensity, it can be easily measured.
This is because one can measure the power of the current
\be
\label{128}
P(t) = R \; j^2(t) \; ,
\ee
which is generated in a coil by the radiating spins. Employing the relation
(14) between the electric current and the induced magnetic field, we get
$$
j^2(t) = \frac{V_c}{4\pi L} \; H^2(t) \; .
$$
The resonator field $H(t)$ can be found from Eq. (57). Neglecting the thermal
Nyquist noise implies $\bt=0$. Averaging over fast oscillations yields
$$
\gm_n^2\; \overline{H^2(t)} = 2\al^2(t) w(t) \; .
$$
Using this, for the averaged current power (128), we have
\be
\label{129}
\overline P(t) =  g \Gm_2 I \hbar \om N \left ( 1 - e^{-\gm t}\right )^2 \;
w(t) \; .
\ee
Comparing Eqs. (129) and (127) for $\gm t\gg 1$, we find
\be
\label{130}
\frac{\overline P(t)}{\overline I(t)} =
\frac{3Q\lbd^3}{8\pi^2 V_c} \; ,
\ee
where $\lbd=2\pi c/\om$. The ratio (130) can be quite large. For instance,
under $\om\sim 10^8$ Hz, hence $\lbd\sim 10^3$ cm, $Q\sim 100$, and
$V_c\sim 10$ cm$^3$, one gets the order of $10^8$. This explains why even
a very low radiation intensity can be easily measured, as is mentioned in
Section 2.

\section{Electron-Nuclear Hyperfine Coupling}

Since real condensed matter contains both nuclei as well as electrons, it is
important to understand how the latter could influence the nuclear spin dynamics.
Electron and nuclear spins interact with each other through hyperfine forces.
A combined system of nuclear and electronic spins is described [53] by the
Hamiltonian
\be
\label{131}
\hat H = \hat H_{nuc} + \hat H_{ele} +\hat H_{hyp} \; ,
\ee
consisting of the parts related to nuclei, $\hat H_{nuc}$, electrons,
$\hat H_{ele}$, and their hyperfine interactions, $\hat H_{hyp}$.

The nuclear Hamiltonian
\be
\label{132}
\hat H_{nuc} = \sum_i \hat H_i +\frac{1}{2} \; \sum_{i\neq j} \hat H_{ij}
\ee
is the same as in Eq. (1), with the Zeeman term (2) and the dipolar term (4).
The electronic spin Hamiltonian is
\be
\label{133}
\hat H_{ele} = -\; \frac{1}{2} \; \sum_{i\neq j} \; J_{ij} \;
{\bf S}_i\cdot {\bf S}_j + \hbar \gm_e \sum_i {\bf B}\cdot{\bf S}_i \; ,
\ee
in which $J_{ij}$ is an exchange interaction potential, $\bS_i$ is an electron
spin, and $\gm_e$ is the electron gyromagnetic ratio defining the electron
magnetic moment $\mu_e=\hbar\gm_e S=0.928\times 10^{-20}$ erg G$^{-1}$, which
approximately equals the Bohr magneton $\mu_B=0.927\times10^{-20}$ erg G$^{-1}$.
The total magnetic field acting on electron spins is the same field (3) as that
acting on nuclear spins. The dipolar interactions between electrons are much
weaker than their exchange interactions.

The Hamiltonian of hyperfine interactions is
\be
\label{134}
\hat H_{hyp} =  A \sum_i \bI_i \cdot\bS_i + \frac{1}{2} \sum_{i\neq j} \;
\sum_{\al\bt} A_{ij}^{\al\bt} \; I_i^\al\; S_j^\bt \; .
\ee
The first term here is the Fermi contact hyperfine interaction between nuclei
and $s$-electrons, with the energy
\be
\label{135}
A= \frac{8\pi}{3}\; \hbar^2\gm_n \gm_e |\psi(0)|^2 \; ,
\ee
where $\psi(\br)$ is the electron wave function. Using the estimate
$|\psi(0)|^2\sim 3/4\pi r_B^3$, where $r_B=5.292\times 10^{-9}$ cm is the Bohr
radius, we have
$$
A\sim \frac{2\hbar^2\gm_n\gm_e}{r_B^3} \; .
$$
Since $\mu_0=\hbar\gm_n$, for protons, with $\gm_n=2.675\times 10^4$
G$^{-1}$s$^{-1}$ we get $A/\hbar\sim 10^9$ s$^{-1}$. The second term in the
hyperfine Hamiltonian (134) describes the dipolar interactions, with the
dipolar tensor
\be
\label{136}
A_{ij}^{\al\bt} = - \hbar^2\; \frac{\gm_n\gm_e}{r_{ij}^3}\;\left (
\dlt_{\al\bt} - 3 n_{ij}^\al\; n_{ij}^\bt \right ) \; .
\ee
Similarly to Eqs. (6), one has
\be
\label{137}
\sum_{j(\neq i)} A_{ij}^{\al\bt} = 0 \; , \qquad
\sum_\al A_{ij}^{\al\al} = 0 \; .
\ee
It is again convenient to pass to the ladder spin operators $S_j^\pm\equiv
S_j^x\pm i S_j^y$. Then the electron spin Hamiltonian (133) can be written as
\be
\label{138}
\hat H_{ele} =\sum_i \hat H_i^{ele} + \frac{1}{2} \;
\sum_{i\neq j} \hat H_{ij}^{ele} \; ,
\ee
where in the Zeeman term, one has
\be
\label{139}
\hat H_i^{ele} = \hbar \gm_e B_0 S_i^z + \frac{1}{2}\; \hbar \gm_e \left ( B_1 +
H \right ) \left ( S_i^+ + S_i^- \right ) \; ,
\ee
and the interaction term contains
\be
\label{140}
\hat H_{ij}^{ele} = - J_{ij} \left ( S_i^+ S_i^- + S_i^z S_j^z \right ) \; .
\ee
The hyperfine Hamiltonian (134) can be presented in the form
\be
\label{141}
\hat H_{hyp} = \sum_i \hat H_i^{hyp} + \frac{1}{2} \sum_{i\neq j}
\hat H_{ij}^{hyp} \; .
\ee
Here the first term is due to contact interactions, with
\be
\label{142}
\hat H_i^{hyp} = \frac{1}{2}\; A \left ( I_i^+ S_i^- + I_i^- S_i^+ +
2I_i^z S_i^z \right ) \; .
\ee
And the second term corresponds to dipolar hyperfine interactions, with
$$
\hat H_{ij}^{hyp} = \al_{ij} \left ( I_i^z S_j^z - \; \frac{1}{4}\;
I_i^+ S_j^- - \; \frac{1}{4} \; I_i^- S_j^+ \right ) +
\bt_{ij} I_i^+ S_j^+ + \bt_{ij}^* I_i^- S_j^- +
$$
\be
\label{143}
+
\sigma_{ij} \left ( I_i^+ S_j^z + I_i^z S_j^+ \right ) +
\sigma_{ij}^* \left ( I_i^z S_j^- + I_i^- S_j^z \right ) \; ,
\ee
where we use the notation
\be
\label{144}
\al_{ij} \equiv A_{ij}^{zz} \; , \qquad \sgm_{ij}\equiv \frac{1}{2}\left (
A_{ij}^{xz} - i A_{ij}^{yz} \right ) \; , \qquad
\bt_{ij} \equiv \frac{1}{4} \left ( A_{ij}^{xx} - A_{ij}^{yy} - 2i\;
A_{ij}^{xy} \right ) \; .
\ee
From the first of Eqs. (137), we have
$$
\sum_{j(\neq i)} \al_{ij} = \sum_{j(\neq i)} \bt_{ij} = \sum_{j(\neq i)}
\sgm_{ij} = 0 \; .
$$
The resonator feedback field is, as early, defined by the Kirchhoff equation
(18), but with
\be
\label{145}
m_x = \frac{\hbar}{V} \; \sum_i \left ( \gm_n <I_i^x>\; -
\gm_e <S_i^x>\right ) \; .
\ee
With this magnetization density, the integral presentation (20) is again valid.

Writing down the Heisenberg equations of motion, we employ the commutation
relations for the nuclear spin operators $I_i^\al$ and also for electron spins,
$$
[S_i^+,S_j^-] = 2\dlt_{ij} S_i^z \; , \qquad [S_i^z,S_j^\pm] = \pm \dlt_{ij}
S_i^\pm \; , \qquad [ I_i^\al, S_j^\bt] = 0 \; .
$$
Generalizing Eqs. (21), we define the local fields acting on nuclear spins
$$
\xi_0 \equiv \frac{1}{\hbar} \; \sum_{j(\neq i)} \left ( c_{ij} I_j^+ +
c_{ij}^* I_j^- + a_{ij} I_j^z + \frac{1}{2}\; \sgm_{ij} S_j^+ +
\frac{1}{2}\; \sgm_{ij}^* S_j^- + \frac{1}{2}\; \al_{ij} S_j^z
\right ) \; ,
$$
\be
\label{146}
\xi \equiv \frac{i}{\hbar} \; \sum_{j(\neq i)} \left ( 2b_{ij} I_j^+
- \; \frac{1}{2}\; a_{ij} I_j^- + 2c_{ij} I_j^z + \bt_{ij} S_j^+ -\;
\frac{1}{4}\; \al_{ij} S_j^- + \sgm_{ij} S_j^z \right ) \; ,
\ee
and local fields acting on electron spins,
$$
\vp_0 \equiv \frac{1}{2\hbar} \; \sum_{j(\neq i)} \left ( \sgm_{ij} I_j^+
+ \sgm_{ij}^* I_j^- +\al_{ij} I_j^z\right ) + \frac{1}{\hbar}\; \left (
J_0 S_i^z -\sum_{j(\neq i)} J_{ij} S_j^z \right ) \; ,
$$
\be
\label{147}
\vp \equiv \frac{i}{\hbar} \; \sum_{j(\neq i)} \left ( \bt_{ij} I_j^+
-\; \frac{1}{4}\; \al_{ij} I_j^- + \sgm_{ij} I_j^z\right ) +
\frac{1}{\hbar}\; \left ( J_0 S_i^- -\sum_{j(\neq i)} J_{ij} S_j^-
\right ) \; ,
\ee
where $J_0\equiv \sum_{j(\neq i)} J_{ij}$. Introduce the frequency operators
\be
\label{148}
\hat\Om_n \equiv -\gm_n B_0 + \frac{A}{\hbar}\; S_i^z + \xi_0 \; , \qquad
\hat\Om_e \equiv \gm_e B_0 + \frac{A}{\hbar}\; I_i^z + \vp_0 \; ,
\ee
and the force operators
\be
\label{149}
\hat f_n \equiv - i\gm_n (B_1 +H) + i\; \frac{A}{\hbar}\; S_i^- + \xi \; ,
\qquad \hat f_e \equiv i\gm_e (B_1 +H) + i\; \frac{A}{\hbar}\; I_i^- +
\vp_0 \; .
\ee
Then the Heisenberg equations of motion can be presented in the form
$$
\frac{dI_i^-}{dt} = - i\hat\Om_n I_i^- + I_i^z\hat f_n \; , \qquad
\frac{dI_i^z}{dt} = - \; \frac{1}{2} \left (\hat f_n^+ I_i^- + I_i^+
\hat f_n\right ) \;
,
$$
\be
\label{150}
\frac{dS_i^-}{dt} = - i\hat\Om_e S_i^- + S_i^z\hat f_e \; , \qquad
\frac{dS_i^z}{dt} = - \; \frac{1}{2} \left (\hat f_e^+ S_i^- + S_i^+
\hat f_e\right ) \;
.
\ee

In the stationary regime, Eqs. (150) define coupled nuclear-electron spin
waves, whose description can be done similarly to Section 4. To accomplish
this, we may define, by analogy with (24),
\be
\label{151}
S_i^\al \equiv \; < S_i^\al>\; + \dlt S_i^\al \; .
\ee
Here, as in Eq. (25), we assume an ideal lattice, for which
$<S_i^\al>=<S_j^\al>$. Considering the case without transverse magnetic
fields, when $B_1=H=0$, we set $<S_i^+>=0$, while $<S_i^z>\not\equiv 0$. This
is analogous to setting $<I_i^\pm>=0$, while $<I_i^z>\not\equiv 0$. Hence,
$I_i^\pm=\dlt I_i^\pm$ and $S_i^\pm=\dlt S_i^\pm$, because of which the local
fields (146) and (147) correspond to random local spin fluctuations, with
$\xi_0=\dlt\xi_0$, $\xi=\dlt\xi$, $\vp_0=\dlt\vp_0$, and $\vp=\dlt\vp$. Then
the second and fourth of Eqs. (150) give $\dlt I_i^z=0$ and $\dlt S_i^z=0$.
To the remaining equations, we substitute the Fourier transformed nuclear spin
operators (28) and
\be
\label{152}
S_j^\pm = \sum_k \; S_k^\pm\; \exp\left (\mp i\bk\cdot\br_j\right ) \; .
\ee
We invoke the Fourier transforms (29) and also introduce
$$
\al_k \equiv \sum_{j(\neq i)} \al_{ij} \; \exp\left ( -i\bk \cdot
\br_{ij} \right ) \; , \qquad \bt_k \equiv \sum_{j(\neq i)} \bt_{ij} \;
\exp\left ( -i\bk \cdot \br_{ij} \right ) \; ,
$$
\be
\label{153}
J_k \equiv \sum_{j(\neq i)} J_{ij} \; \exp\left ( -i\bk \cdot \br_{ij}
\right ) \; .
\ee
Generalizing Eqs. (31), we use the notation
$$
\mu_k \equiv -\gm_n B_0 + \frac{a_k}{2\hbar}\; < I_i^z> \; +
\frac{A}{\hbar}\; < S_i^z> \; , \qquad
\lbd_k \equiv \frac{2b_k}{\hbar}\; <I_i^z> \; ,
$$
\be
\label{154}
\ep_k \equiv \gm_e B_0 + \frac{1}{\hbar}\; ( J_k - J_0 ) \;
< S_j^z> \; + \frac{A}{\hbar}\; < I_j^z> \; .
\ee
Then Eqs. (150) can be reduced to the equations
$$
i\; \frac{dI_k^-}{dt} = \mu_k I_k^- - \lbd_k I_k^+ - \;
\frac{1}{\hbar}\left ( A - \; \frac{\al_k}{4}\right ) \; < I_i^z>\;
S_k^- - \; \frac{\bt_k}{\hbar} \; < I_i^z> \; S_k^+ \; ,
$$
\be
\label{155}
i\; \frac{dS_k^-}{dt} = \ep_k S_k^- - \; \frac{1}{\hbar} \left ( A - \;
\frac{\al_k}{4}\right ) \; <S_i^z>\; I_k^- - \; \frac{\bt_k}{\hbar} \;
< S_i^z>\; I_k^+ \; ,
\ee
which describe coupled nuclear-electron spin waves. The spectrum of the
latter follows from Eqs. (155) yielding the fourth-order algebraic equation
defining two positive branches of coupled spin-wave collective excitations.
The coupling between the branches comes from the terms of Eqs. (155)
containing the hyperfine parameters $A$, $\al_k$, and $\bt_k$. Without
the latter, there would exist two separate types of collective excitations
corresponding to nuclear spin waves with the spectrum (33) and to electron
spin waves with the spectrum $\ep_k$. In the presence of the hyperfine
coupling, one may conditionally distinguish one of the branches of the
coupled collective excitations as the branch related to nuclear spin waves,
while another branch can be ascribed to electron spin waves. The hyperfine
coupling, of course, essentially modifies the spectra of the coupled
nuclear-electron spin waves. In particular, nuclear spin waves become
noticeably influenced by the indirect exchange through electron spins,
when the latter possess a long-range magnetic order. The related exchange
force is called the Suhl-Nakamura force [54], which can stabilize the
nuclear spin waves. When nuclear spins are initially in a nonequilibrium
state, nuclear-electron spin waves play the role of a trigger starting the
motion of nuclear spins from the nonequilibrium to their equilibrium state.

\section{Enhanced Nuclear Radiation}

To describe the nonequilibrium dynamics of coupled nuclear and electron
spins, we again employ the scale separation approach [9,42,56,57,106].
To this end, the variables (146) and (147), characterizing local field
fluctuations, are treated as random fields. The stochastic averages for
the random variables (146) are defined as in Eq. (41). However, the
dynamic broadening $\Gm_3$, in the presence of hyperfine interactions,
is renormalized according to the relation
\be
\label{156}
\Gm_3^2 = \Gm_{nn}^2 + \Gm_{ne}^2 \; ,
\ee
where $\Gm_{nn}\approx\rho_n\gm_n\mu_n$ is the broadening due to dipolar
nuclear interactions, with $\rho_n$ being the density of nuclear spins, and
$\Gm_{ne}\approx\sqrt{\rho_n\rho_e}\gm_n\mu_e$ is the broadening caused by
the hyperfine nuclear-electron interactions, with $\rho_e$ being the density
of electron spins.

For the random fields (147), the stochastic averages can be defined as
$$
\ll \vp_0(t)\gg \; = \; \ll \vp(t)\gg \; = 0 \; , \qquad
\ll \vp_0(t)\vp_0(t')\gg \; = 2\gm_3 \dlt(t-t') \; ,
$$
\be
\label{157}
\ll \vp_0(t)\vp(t')\gg \; = \; \ll \vp(t)\vp(t')\gg \; = 0 \; , \qquad
\ll \vp^*(t)\vp(t')\gg \; = 2\gm_3 \dlt(t-t') \; ,
\ee
with the dynamic broadening $\gm_3$ given by the relation
\be
\label{158}
\gm_3^2 = \Gm_{ee}^2 + \Gm_{en}^2 \; .
\ee
Here, $\Gm_{ee}\approx\rho_e\gm_e\mu_e$ is the broadening caused by electron
spin fluctuations, while $\Gm_{en}\approx\sqrt{\rho_e\rho_n}\gm_e\mu_n$ is
the broadening due to electron-nuclear hyperfine interactions. Recall that
$\mu_n\equiv\hbar\gm_n I$ and $\mu_e\equiv\hbar\gm_e S$ are the nuclear and
electron magnetic moments, respectively.

To estimate the related broadening widths, we may assume that $\rho_n\sim
\rho_e\sim 10^{23}$ cm$^{-3}$. Then, with $\gm_n=2.675\times 10^4$
G$^{-1}$s$^{-1}$, $\gm_e=1.759\times 10^7$ G$^{-1}$s$^{-1}$, $\mu_n=1.411
\times 10^{-23}$ erg G$^{-1}$, and $\mu_e=0.928\times 10^{-20}$ erg G$^{-1}$,
we find $\Gm_{nn}\sim 10^4$ s$^{-1}$, $\Gm_{ne}\sim 10^7$ s$^{-1}$,
$\Gm_{ee}\sim 10^{10}$ s$^{-1}$, and $\Gm_{en}\sim
10^7$ s$^{-1}$. As is seen,
hyperfine interactions may strongly influence
local nuclear spin fluctuations
but are not important for electron spin motion.

By analogy with Eqs. (43) to (45), we introduce the functional variables for
electron spins: the transition function
\be
\label{159}
x \equiv \frac{1}{SN_e} \; \sum_{j=1}^{N_e} < S_j^-> \; ,
\ee
where $N_e$ is the number of electrons, the coherence intensity
\be
\label{160}
y \equiv \frac{1}{S^2N_e(N_e-1)} \; \sum_{i\neq j}^{N_e} \;
<S_i^+ S_j^-> \; ,
\ee
and the electron spin polarization
\be
\label{161}
z \equiv \frac{1}{SN_e} \; \sum_{j=1}^{N_e} \; <S_j^z> \; .
\ee
The averaging in Eqs. (159) to (161) is accomplished over the spin degrees of
freedom, not involving the random variables (146) and (147). The stochastic
averages for the latter are defined in Eqs. (41) and (157). The random variables
ascribed to nuclear and electron fluctuations are assumed to be uncorrelated
with each other.

Employing expressions (159) to (161), we may present the averages of the
frequency operators (148) as
\be
\label{162}
<\hat\Om_n> \; = \om_n + \xi_0 \; , \qquad
<\hat\Om_e> \; = \om_e + \vp_0 \; ,
\ee
where the effective nuclear and electron frequencies are
\be
\label{163}
\om_n \equiv -\gm_n B_0 + \frac{1}{\hbar}\; A S z \; , \qquad
\om_e \equiv \gm_e B_0 + \frac{1}{\hbar}\; A I s \; .
\ee
For an external field $B_0\sim 1$ T$=10^4$ G, since $\gm_n\sim 10^4$
G$^{-1}$s$^{-1}$, $\gm_e\sim 10^7$ G$^{-1}$s$^{-1}$, and $A/\hbar\sim 10^9$
s$^{-1}$, we see that the hyperfine term in the nuclear frequency $\om_n$
can become larger than the nuclear Zeeman frequency $\om_0\equiv|\gm_nB_0|
\sim 10^8$ Hz, provided that there exists a nonzero average electron
magnetization, with $z\neq 0$. Hence, it is necessary to accurately take
into account the possible hyperfine shift of the nuclear frequency. At the
same time, the electron Zeeman frequency $\gm_eB_0\sim 10^{11}$ Hz is much
larger than the hyperfine frequency shift, so that $\om_e\approx\gm_eB_0$.

For the averages
\be
\label{164}
f_n \equiv \; < \hat f_n>\; , \qquad f_e \equiv \; < \hat f_e>
\ee
of the force operators (149), we have
\be
\label{165}
f_n = -i\gm_n (B_1+H) + \frac{i}{\hbar}\; AS x + \xi \; , \qquad
f_e = i\gm_e (B_1+H) + \frac{i}{\hbar}\; AI u + \vp \; .
\ee
Here, the hyperfine terms play the role of additional fields acting on nuclear
and electron spins.

It is worth mentioning that the hyperfine shifts in the effective frequencies
(163) are caused by the first-order hyperfine effect. Generally, there exists
the second-order effect due to the Suhl-Nakamura forces. The second-order
frequency shift appears as follows. Let us consider the evolution equations
(150) for the electron spin operators, with the explicit form of the variables
(147), without treating the latter as random. Then, the third of Eqs. (150)
yields
$$
S_i^-(t) \approx \left [ S_i^-(0) + i\; \frac{\hat F_{en}}{\om_e} \;
S_i^z \right ] e^{-i\om_e t} - i\; \frac{\hat F_{en}}{\om_e}\; S_i^z \; ,
$$
where
$$
\hat F_{en} \equiv \frac{i}{\hbar} \sum_{j(\neq i)} \left ( \bt_{ij} I_j^+ - \;
\frac{1}{4}\; \al_{ij} I_j^- + \sgm_{ij} I_j^z \right ) \; .
$$
Substituting this to the first of equations (150), we come to the conclusion
that the effective nuclear frequency should have the form
\be
\label{166}
\om_n = \om_0 + \frac{A}{\hbar}\; < S_i^z> + \Dlt_{SN} \; ,
\ee
in which the Suhl-Nakamura frequency shift is
\be
\label{167}
\Dlt_{SN} = \sum_{j(\neq i)} \frac{|\sgm_{ij}|^2}{\hbar^2\om_e} \;
<S_i^z>\; s \; .
\ee
This is often called the dynamic frequency shift because of its dependence on
$s=s(t)$.

However, the second-order frequency shift (167) is much smaller than the
first-order hyperfine shift in Eq. (166),
$$
\left | \frac{\hbar\Dlt_{SN}}{A} \right | \ll 1 \; .
$$
Thus, Eq. (167) shows that $\Dlt_{SN}\sim n_0S\rho_n\rho_e(\hbar\gm_n\gm_e)^2/
\om_e$, provided that there is a long-range magnetic order of electron spins,
i.e., $<S_i^z>\neq 0$. Here $n_0\sim 10$ is the number of nearest neighbors.
For the typical values of parameters, this gives $\Dlt_{SN}\sim 10^4$ Hz,
which is much smaller than $A/\hbar\sim 10^9$ Hz, hence $\hbar\Dlt_{SN}/A
\sim 10^{-5}$ is really a very small value. This makes it admissible to take
account of only first-order hyperfine frequency shifts.

Averaging Eqs. (150) over the spin degrees of freedom, we shall use the
decoupling of nuclear and spin operators
\be
\label{168}
<I_i^\al S_j^\bt>\; = \; < I_i^\al><S_j^\bt> \; ,
\ee
valid for all $i$ and $j$. And, similarly to Eq. (42), we employ the decoupling
\be
\label{169}
<S_i^\al S_j^\bt> \; = \; <S_i^\al><S_j^\bt> \qquad (i\neq j)
\ee
for different indices $i$ and $j$. Recall that this type of decoupling has been
called the stochastic mean-field approximation, since the averages $<\ldots>$ do
not include the averaging over the stochastic degrees of freedom, thus, allowing
for the consideration of quantum effects [8,9].

Finally, from Eqs. (150), we come to the stochastic differential equations
describing the evolution of nuclear spins,
$$
\frac{du}{dt} = - i ( \om_n +\xi_0 - i\Gm_2) u + f_n s\; , \qquad
\frac{dw}{dt} = -2\Gm_2 w + \left ( u^* f_n + f_n^* u \right ) s \; ,
$$
\be
\label{170}
\frac{ds}{dt} = -\; \frac{1}{2}\left ( u^* f_n + f_n^* u \right ) -
\Gm_1 (s -\zeta) \; ,
\ee
as well as the motion of electron spins,
$$
\frac{dx}{dt} = - i(\om_e +\vp_0 - i\gm_2 ) x  + f_e z \; , \qquad
\frac{dy}{dt} = - 2\gm_2 y + ( x^* f_e + f_e^* x) z \; ,
$$
\be
\label{171}
\frac{dz}{dt} = -\; \frac{1}{2}\left ( x^* f_e + f_e^* x\right ) -
\gm_1 (z -\sgm) \; ,
\ee
where $\gm_1$ and $\gm_2$ are the spin-lattice and spin-spin relaxation
parameters for electronic spins, while $\sgm$ is an equilibrium value of
$<S_i^z>/S$. Equations (170) and (171) are to be complimented by the Kirchhoff
equation (18), with $m_x$ given by Eq. (145), which reads
\be
\label{172}
m_x = \frac{1}{2}\; \rho_n\mu_m \left ( u^* + u\right ) -\; \frac{1}{2}\;
\rho_e\mu_e \left ( x^* + x\right ) \; .
\ee

The evolution equations for nuclear and electron spins are very similar to each
other. This means that we could consider electron spin superradiance on the
absolutely same footing as the nuclear spin superradiance. The major difference
is that hyperfine interactions can essentially influence the motion of nuclear
spins, while these interactions do not disturb much the electron spin motion.
Therefore, we concentrate now our attention on the evolution of nuclear spins.

The system of equations (170) plus (171) can again be solved involving the scale
separation approach [8,9,42,56,57]. For this purpose, we remember the existence
of several small parameters defined in Eqs. (49), (51), (55), and also assume
the validity of the resonance condition (56). In addition, we have as well the
inequalities
\be
\label{173}
\frac{\gm_1}{|\om_e|} \ll 1 \; , \qquad \frac{\gm_2}{|\om_e|} \ll 1 \; ,
\qquad
\frac{\gm_3}{|\om_e|} \ll 1 \; .
\ee
Also, we have to keep in mind that
\be
\label{174}
\frac{\mu_n}{\mu_e} = \frac{\gm_n I}{\gm_e S} \ll 1 \; .
\ee

Being based on these inequalities, we may realize the following {\it
classification of relative quasi-invariants}:  The functions $w$ and $s$ are
temporal quasi-invariants with respect to the functional variables $u,x,y$,
and $z$; the functions $y$ and $z$ are quasi-invariants with respect to $x$;
and $u$ is also a quasi-invariant with respect to $x$.

To derive the expression for the resonator feedback field, we again use the
integral form (20), iterating it with the approximate solutions for $x$ and
$u$, keeping in mind that, because of the inequalities (174), we have
$|\om_n/\om_e|\ll 1$. We substitute in integral (20) the solutions
$$
x \cong \left ( x_0 -\; \frac{AIz}{\hbar\om_e}\; u \right )
e^{-i\om_e t} + \frac{AIz}{\hbar\om_e}\; u
$$
and $u\cong u_0\exp(-i\om_n t)$. This shows that electron spins oscillate with
the frequencies $\om_e$, $\om_n$, and $\om_e+\om_n$. Calculating the transverse
magnetization (172), we insert it into integral (20), replacing $z$ by its average
$\sgm$. In that way, we come to the expression $\gm_nH=\mu_0H/\hbar$ having the
same form (57), but with the coupling function
$$
\tilde\al =\tilde g \Gm_2 \left ( 1 - e^{-\gm t}\right ) \; ,
$$
instead of $\al$, where the renormalized spin-resonator coupling is
$$
\tilde g\equiv g \left ( 1 - \;
\frac{\rho_e\mu_e AI\sgm}{\rho_n\mu_n\hbar\om_e} \right ) \; .
$$
Here $g$ is defined in Eq. (60) and $\om_e\cong\gm_eB_0$. Taking into account
condition (46), according to which $\gm_nB_0<0$, the renormalized spin-resonator
coupling can be presented as
\be
\label{175}
\tilde g\equiv g \left ( 1 +
\frac{\rho_eAS\sgm}{\rho_n\hbar\om_0} \right ) \; ,
\ee
where $\om_0\equiv|\gm_nB_0|$. Analogously to the previous procedures, we
may study different regimes of coherent nuclear radiation, just replacing
everywhere the spin resonator coupling (60) by its renormalized value (175)
and by using expression (156) for the dynamic broadening $\Gm_3=
\sqrt{\Gm_{nn}^2+\Gm_{ne}^2}$.

The noticeable renormalization of the coupling (175) occurs when there exists
a long-range magnetic order in electron spins, so that $\sgm\neq 0$. Then the
coupling (175) can become an order larger than $g$, that is, $\tilde g/g\sim
10$. This, respectively, leads to the {\it enhancement of nuclear spin
radiation}, with the current power (129) becoming an order stronger.
Simultaneously, the crossover time (98), or (100), and the pulse time (108)
become an order shorter. Thus, in the case of pure spin superradiance, when
$w_0=0$, and for $\tilde gs_0 \gg 1$, we have
$$
t_c \simeq \frac{\tau}{\tilde g s_0} \; , \qquad
\tau_p \simeq \frac{T_2}{\tilde g s_0} \; .
$$
Then the delay time (103) changes to
$$
t_0 \simeq \frac{\tau}{\tilde gs_0} + \frac{T_2}{2\tilde gs_0}\;
\ln\left | \frac{2}{\tilde\Gm_3\tau} \right | \; ,
$$
which shows that $t_0$ also becomes about an order shorter.

It is important to stress that the enhancement effect can exist not only in
ferromagnets or ferrimagnets, but also in paramagnets, provided that the
external field is sufficiently large and the temperature is low. Thus, for
a paramagnetic system of electrons, the average magnetization per spin, due
to the external magnetic field $B_0$, is
$$
\sgm = {\rm tanh} \left ( \frac{\mu_eB_0}{k_BT} \right ) \; .
$$
For the typical values of parameters, with $B_0\sim 10^4$ G, we have $\mu_eB_0
\sim 10^{-16}$ erg, hence $\mu_eB_0/\hbar\sim 10^{11}$ s$^{-1}$. The thermal
frequency $\om_T\equiv k_BT/\hbar$, at $T\sim 0.1$ K, is $\om_T\sim 10^{11}$
s$^{-1}$, which is of the same order as $\mu_eB_0/\hbar\sim\om_e$. Then
$\sgm\sim{\rm tanh}(1)\approx 0.76$. Therefore, the enhancement effect,
under such conditions, exists for a paramagnet as well.

\section{Superradiance by Magnetic Molecules}

There exists an interesting class of composite objects behaving as giant nuclei
with total spins ranging from $1/2$ to rather large values. These are magnetic
molecules [60,107--109], which can form crystalline materials termed molecular
magnets. In these materials, all magnetic clusters possess the same shape, size,
and orientation, because of which the inhomogeneous broadening caused by the
system nonuniformity is very low. As examples of such magnetic molecules, we
may mention the following, where in brackets the total ground-state spin
is shown:
$$
{\rm K}_6\left [ {\rm V}_{15}^{4+}{\rm As}_6{\rm O}_{42}({\rm H}_2{\rm O})
\right ] \cdot 8 {\rm H}_2{\rm O} \qquad \left ( {\rm S} = 1/2 \right )
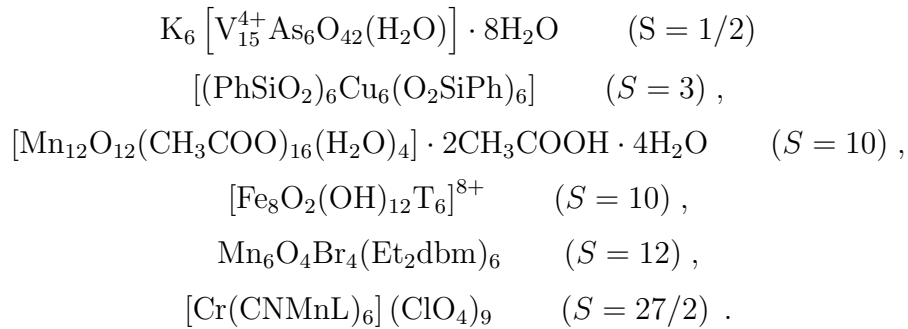
$$
$$
\left [ ({\rm Ph}{\rm Si}{\rm O}_2)_6 {\rm Cu}_6 ({\rm O}_2{\rm Si}{\rm Ph})_6
\right ]  \qquad (S=3) \; ,
$$
$$
\left [ {\rm Mn}_{12}{\rm O}_{12}({\rm CH}_3{\rm COO})_{16}({\rm H}_2{\rm O})_4
\right ] \cdot 2{\rm C}{\rm H}_3{\rm COOH} \cdot 4{\rm H}_2{\rm O}
\qquad (S=10) \; ,
$$
$$
\left [ {\rm Fe}_8{\rm O}_2({\rm OH})_{12}{\rm T}_6 \right ]^{8+}
\qquad (S=10) \; ,
$$
$$
{\rm Mn}_6{\rm O}_4{\rm Br}_4 ({\rm Et}_2{\rm dbm})_6 \qquad (S=12) \; ,
$$
$$
\left [ {\rm Cr}({\rm CNMnL})_6\right ] ({\rm ClO}_4)_9 \qquad
\left ( S = 27/2 \right ) \; .
$$

The first of the above molecules is briefly denoted as $V_{15}$. Molecular
magnets formed by these molecules have no magnetic anisotropy. The short-hand
notation for the third molecule is Mn$_{12}$. The corresponding molecular
magnets possess a rather strong single-site anisotropy characterized by the
parameter $D\approx 0.967\times 10^{-16}$ erg, which gives $D/k_B\approx 0.7$
K and $D/\hbar\approx 0.917\times 10^{11}$ s$^{-1}$. At temperatures below the
blocking temperature $T_B\approx 3$ K, the magnetization of a molecular crystal
formed of  Mn$_{12}$ is preserved during the relaxation time $T_1\sim 10^7$ s.
The size of each molecule Mn$_{12}$ is about $10$ $\AA$ and the distance
between the nearest neighbours is around $14$ $\AA$. Hence the average density
is $\rho\approx 0.364\times 10^{21}$ cm$^{-3}$. The spin-spin relaxation
parameter is $\Gm_2=n_0\rho\gm_S|\mu_0|S$, where
\be
\label{176}
\mu_0 = - g_S\mu_B = -\hbar\gm_S \; ,
\ee
$g_S$ is the Land\'e factor; $\mu_B$, Bohr magneton, and $\gm_S$ is the
gyromagnetic ratio of a molecule with spin $S$. This parameter $\Gm_2\sim
10^{10}$ s$^{-1}$ is due to dipolar spin interactions.

The fourth molecule has the abbreviation Fe$_8$. In its chemical formula, the
letter T stands for the organic ligand triazacyclononane. The corresponding
molecular magnet possesses the magnetic anisotropy described by the parameter
$D\approx 0.414\times 10^{-16}$ erg. Then, $D/k_B\approx 0.3$ K and $D/\hbar
\approx 0.392\times 10^{11}$ s$^{-1}$. Below the blocking temperature $T_B
\approx 1$ K, the relaxation of magnetization to zero in a molecular magnet
occurs during the relaxation time $T_1\sim 10^5$ s. The average density is
$\rho\approx 0.4\times 10^{21}$ cm$^{-3}$. The interaction between molecules
is also through dipolar forces, resulting in the spin-spin attenuation
$\Gm_2\sim 10^{10}$ s$^{-1}$.

In the last molecule from the list above, the letter L in its chemical formula
means a neutral pentadentate ligand. Its properties are close to those of the
molecules Mn$_{12}$ and Fe$_8$ described above.

In principle, at very low temperatures, of the order of $\hbar\Gm_2/k_B\sim
0.1$ K, a purely dipolarly interacting molecular magnet can become
magnetically ordered. The paramagnet-ferromagnet transition at $T_c\approx
0.16$ K was observed [110] in the dipolar magnet formed of the molecules of
the fifth type from the list above, this kind of a molecule being abbreviated
as Mn$_6$. The molecular magnet composed of the latter molecules has a week
anisotropy, with $D\approx 1.795\times 10^{-18}$ erg. Thus, $D/k_B\approx
0.013$ K and $D/\hbar\approx 1.7\times 10^9$ s$^{-1}$. The spin-spin dipolar
relaxation parameter is $\Gm_2\sim 10^{10}$ s$^{-1}$. The spin-lattice
relaxation time, at $B_0\sim 10^4$ G and $T\sim 0.1$ K, is $T_1\sim 10^{-3}$
s, so that $\Gm_1\sim 10^3$ s$^{-1}$. Generally, $T_1$ can be estimated
from the formula
$$
T_1 \approx \tau_0\exp\left ( \frac{\hbar\gm_SB_0}{k_BT} \right ) \; .
$$
For the molecule Mn$_6$, one has $\tau_0\approx 3\times 10^{-4}$ s.

The Hamiltonian of a molecular magnet, consisting of $N$ magnetic molecules,
each having spin $S$, has the form
\be
\label{177}
\hat H = \sum_i \hat H_i + \frac{1}{2} \sum_{i\neq j} \hat H_{ij} \; ,
\ee
in which $\hat H_i$ is related to individual spins and $\hat H_{ij}$, to pair
spin interactions. The individual term
\be
\label{178}
\hat H_i = -\mu_0 \bB\cdot\bS_i - D(S_i^z)^2
\ee
includes the part characterizing the single-site magnetic anisotropy with
the anisotropy parameter $D$. Positive $D>0$ implies an easy-axis anisotropy,
while $D<0$ means an easy-plane anisotropy. The pair term in the Hamiltonian
(177) corresponds to dipolar spin interactions
\be
\label{179}
\hat H_{ij} = \sum_{\al\bt} C_{ij}^{\al\bt} S_i^\al S_j^\bt
\ee
with the dipolar tensor
$$
C_{ij}^{\al\bt} = \frac{\mu_0^2}{r_{ij}^3} \left ( \dlt_{\al\bt} -
3n_{ij}^\al n_{ij}^\bt \right )
\; .
$$
All notations in this sections are close to those of Section 3. The difference
is that now we are considering the molecular spins $S_i^\al$ formed by
electrons. The ground-state molecular spin $S$ can be quite large. The
principal distinction, as compared to Section 3, is the necessity of taking
into account the magnetic anisotropy that can be rather strong. As early,
the dipolar tensor enjoys the properties
$$
\sum_\al C_{ij}^{\al\bt} = 0 \; , \qquad
\sum_{j(\neq i)} C_{ij}^{\al\bt} = 0 \; .
$$
Similar to Eq. (9), we use the notation
$$
a_{ij} \equiv C_{ij}^{zz} \; , \qquad c_{ij} \equiv \frac{1}{2}\left (
C_{ij}^{xz} - i C_{ij}^{yz}\right ) \; , \qquad
b_{ij} \equiv \frac{1}{4}\left ( C_{ij}^{xx} - C_{ij}^{yy} - 2i C_{ij}^{xy}
\right ) \; ,
$$
with Eq. (10) being valid. Employing the ladder spin operators $S_i^\pm\equiv
S_i^x\pm iS_i^y$, we may present the term (178) as
\be
\label{180}
\hat H_i = -\mu_0 B_0 S_i^z - D(S_i^z)^2 - \; \frac{1}{2}\;
\mu_0 (B_1+H) (S_i^+ + S_i^-) \; ,
\ee
where the total magnetic field is assumed to be the same as in Eq. (3). And
the pair term (179) takes the form
$$
\hat H_{ij} = a_{ij}\left ( S_i^z S_j^z -\; \frac{1}{2}\; S_i^+ S_j^-
\right ) +
$$
\be
\label{181}
+ b_{ij} S_i^+ S_j^+ + b_{ij}^* S_i^- S_j^- + 2c_{ij} S_i^+ S_j^z +
2c_{ij}^* S_i^- S_j^z \; .
\ee
In the Kirchhoff equation (18), the magnetization density is
\be
\label{182}
m_x = \frac{\mu_0}{V}\; \sum_i \; < S_i^x> \; .
\ee
The fluctuating local fields are denoted as
$$
\xi_0 \equiv \frac{1}{\hbar} \; \sum_{j(\neq i)} \left ( a_{ij} S_j^z +
c_{ij}^* S_j^- + c_{ij} S_j^+ \right ) \; ,
$$
\be
\label{183}
\xi \equiv \frac{i}{\hbar} \; \sum_{j(\neq i)} \left ( 2c_{ij} S_j^z - \;
\frac{1}{2}\; a_{ij} S_j^- + 2b_{ij} S_j^+ \right ) \; .
\ee
Writing down the Heinsenberg equations for the spin operators, we shall need,
in addition to the standard commutation relations, the relation
$$
\left [ S_i^-, (S_j^z)^2 \right ] = \dlt_{ij} \left ( S_i^- S_i^z +
S_i^z S_i^- \right ) \; .
$$
Thus, the evolution equations for the spin operators, with the notation
\be
\label{184}
f \equiv i\gm_S (B_1 + H) + \xi \; ,
\ee
become
$$
\frac{dS_i^-}{dt} = - i(\gm_S B_0 +\xi_0) S_i^- + fS_i^z +
i\; \frac{D}{\hbar} \left ( S_i^- S_i^z + S_i^z S_i^-\right ) \; ,
$$
\be
\label{185}
\frac{dS_i^z}{dt} = -\; \frac{1}{2}\left ( f^* S_i^- + S_i^+ f\right ) \; .
\ee

Analogously to Section 4, we can show that there exist {\it molecular spin
waves}. A subtle point here is the linearization of the last term in the
first of Eqs. (185), which has to be done so that to take into account the
absence of this term for $S=1/2$. This can be achieved by invoking the
linearization
\be
\label{186}
S_i^- S_i^z + S_i^z S_i^- = \left ( 2 -\; \frac{1}{S}\right ) <S_i^z>
\; S_i^-
\ee
possessing the correct asymptotic behaviour for $S=1/2$ as well as for
$S\ra\infty$ (see discussion in Refs. [60,94]). Introduce the effective
frequency
\be
\label{187}
\om_D \equiv \gm_S B_0  -\; \frac{D}{\hbar} \left ( 2 - \;
\frac{1}{S} \right ) <S_i^z> \; .
\ee
From Eqs. (185), we obtain
\be
\label{188}
\frac{dS_k^-}{dt} = - i\mu_k S_k^- + i\lbd_k S_k^+ \; ,
\ee
where
\be
\label{189}
\mu_k \equiv \frac{a_k}{2\hbar}\; < S_i^z> \; + \om_D \; , \qquad
\lbd_k \equiv \frac{2b_k}{\hbar} \; < S_i^z> \; .
\ee
Then the spectrum of molecular spin waves is described by Eq. (33).
The condition for the spectrum to be positive, in the case of a cubic
symmetry reads
$$
\frac{2\rho\mu_0^2S\om_0}{\hbar\om_D^2} \; < 1 \qquad \left ( \om_0
\equiv \frac{1}{\hbar} |\mu_0 B_0|\right ) \; .
$$
The external longitudinal magnetic field is directed so that, similarly
to the inequality (46), we have
\be
\label{190}
\mu_0 B_0 < 0 \; , \qquad \gm_S B_0 > 0 \; .
\ee

Following Section 5, we define the transition function
\be
\label{191}
u \equiv \frac{1}{SN} \sum_{i=1}^N < S_i^-> \; ,
\ee
the coherence intensity
\be
\label{192}
w \equiv \frac{1}{S^2N(N-1)} \; \sum_{i\neq j}^N < S_i^+ S_j^-> \; ,
\ee
and the spin polarization
\be
\label{193}
s \equiv \frac{1}{SN} \sum_{i=1}^N < S_i^z> \; ,
\ee
where the angle brackets denote the averaging over the spin degrees
of freedom, interpreting the local fields (183) as random variables
with the stochastic averages (41). For the products of spin operators
with different indices, we employ the stochastic mean-field approximation
(42), while for the operators with coinciding indices, we use the
decoupling
$$
<S_i^- S_i^z + S_i^z S_i^-> \; = \left ( 2 -\;
\frac{1}{S} \right ) <S_i^-><S_i^z>
$$
resulting from Eq. (186).

The uniform molecular spin-wave frequency (187) becomes
\be
\label{194}
\om_D = \om_0 - (2S -1) \frac{D}{\hbar}\; s \; ,
\ee
with $\om_0\equiv\gm_SB_0$. Then for the functions (191) to (193) we derive
the same equations (47), except that the first of these equations appears
as
\be
\label{195}
\frac{du}{dt} = - i ( \om_D +\xi_0 -i\Gm_2) u + fs \; ,
\ee
with $\om_0$ replaced by $\om_D$. Note that $\om_0$ is always positive,
while $\om_D$, given by Eq. (194), can be both positive as well as negative.

To deduce the expression for the resonator feedback field, we resort to the
integral Eq. (20). In analogy with Eq. (57), we find
$$
\frac{\mu_0H}{\hbar} = i(\al_D u - \al_D^* u^*) + 2\bt\cos\om t \; ,
$$
where, instead of Eq. (58), we have the coupling function
\be
\label{196}
\al_D = \Gm_0\om_D\left [
\frac{1-\exp\{ -i(\om-\om_D)t-\gm t\} }{\gm+i(\om-\om_D)} +
\frac{1-\exp\{ i(\om+\om_D)t-\gm t\} }{\gm-i(\om+\om_D)} \right ] \; ,
\ee
in which the natural width is
\be
\label{197}
\Gm_0 \equiv \pi\eta\rho\gm_S\mu_S \qquad
(\mu_S \equiv \hbar\gm_S S) \; .
\ee
An efficient coupling with the resonator occurs only when $\om\approx\om_D$,
if $\om_D>0$, or when $\om\approx -\om_D$, if $\om_D<0$. This implies the
resonance condition
$$
\frac{|\Dlt_D|}{\om} \ll 1 \; , \qquad \Dlt_D\equiv \om -|\om_D| \; .
$$
If the resonance is sharp, such that $|\Dlt_D|<\gm$, then the coupling
function (196) reduces to
\be
\label{198}
\al_D = g_D \Gm_2 \left ( 1  - e^{-\gm t} \right ) \; ,
\ee
where the spin-resonator coupling is
\be
\label{199}
g_D \equiv \frac{\gm\Gm_0\om_D}{\Gm_2(\gm^2+\Dlt_D^2)} \; .
\ee

In this way, introducing the effective force
$$
f_1 \equiv -i\nu_0 - i(\nu_1+\bt) e^{-i\om t} + \xi \; ,
$$
we come to the evolution equations
$$
\frac{du}{dt} = - i (\om_D +\xi_0) u - (\Gm_2 -\al_D s) u + f_1 s \; ,
\qquad \frac{dw}{dt} = -2(\Gm_2 -\al_D s) w +\left (
u^* f_1 + f_1^* u  \right ) s \; ,
$$
\be
\label{200}
\frac{ds}{dt} = - \al_D w - \; \frac{1}{2} \left ( u^* f_1 + f_1^* u
\right ) - \Gm_1 (s -\zeta) \; .
\ee
Here we have omitted the nonresonant terms, corresponding to the last
terms of Eqs. (64), which do not contribute to the equations for the
guiding centers in the averaging technique. Using this technique, we
obtain the guiding-center equations of the same form as Eqs. (70), but
with $\al$, $\om_0$, and $\Dlt$ replaced by $\al_D$, $\om_D$, and
$\Dlt_D$, respectively. The following description of possible regimes
of coherent radiation by magnetic molecules is the same as that for
nuclear spins.

First of all, it is necessary to stress that the electromagnetic interaction
of radiating magnetic dipoles through the common radiation field can never
produce coherent radiation. This was explained in detail in Section 6.1,
whose results are valid for arbitrary spins, whether these are nuclear or
molecular spins. The impossibility of collectivizing the spin motion by
means of the magnetodipole radiation is caused by the fact that the
corresponding radiation time $T_{rad}\gg T_2$ is many orders larger than
the spin-spin dephasing time due to dipolar spin interactions. This is
expressed by inequality (85). Therefore, the assumption [111] that molecular
nanomagnets could exhibit superradiance caused by magnetodipole radiation
is wrong. In addition, the regime considered in Ref. [111] requires a very
strong transverse magnetic field of about $6$T$\sim 10^5$ G. Such a strong
transverse field suppresses the single-site anisotropy, whose effective field
is $B_D\equiv(2S-1)D/\hbar\gm_S$. The latter, e.g., for Mn$_{12}$, with $\gm_S
\sim 10^7$ G$^{-1}$s$^{-1}$ and $D/\hbar\sim 10^{11}$ s$^{-1}$, is of the
order of $B_D\sim 10^5$ G. Any magnetic relaxation in the presence of such
a strong transverse magnetic field has nothing to do with superradiance,
which is a self-organized spontaneous radiation. When a transverse field
induces spin relaxation, this is called spin induction. Even if a kind
of spin induction could be realized for magnetic molecules, the related
relaxation time would be of the order of $T_2$. So, it would be just
a simple spin induction, without any collective effects.

In order to realize the real spin superradiance by magnetic molecules,
it is necessary to couple them to a resonant electric circuit, whose
feedback field could collectivize the spin motion. When there are no
transverse external fields, as it should be for the regime of pure
superradiance, the process is triggered by molecular spin waves. Then
the arising feedback field can make the spin motion coherent, in the
same way as it has been considered above for nuclear spins. The
difference with the latter is the presence of magnetic anisotropy
that may complicate the experimental realization of molecular spin
superradiance.

From the evolution equations (200) it is seen that the increase of
the transverse coherence occurs when the effective attenuation
$\Gm_2-g_Ds$ becomes negative, which requires that the spin-resonator
coupling (199) be positive and sufficiently large. To have positive
$g_D$, one needs that the effective frequency (194) be also positive,
that is
$$
\om_0 > (2S-1) \; \frac{D}{\hbar}\; s \; .
$$
This happens, assuming that $s>0$, when $D$ is negative, so that the
molecules form an easy-plane nanomagnet, or, if $D>0$, when the applied
longitudinal field $B_0$ is sufficiently strong. For instance, in the
case of Fe$_8$, the anisotropy field is $B_D\sim 10^4$ G. Hence, the
external field has to be stronger than 1 T.

Another complication is that the frequency (194) is, actually,
a function of time through $s=s(t)$. Therefore, to organize the
resonance condition $\om\approx\om_D$, one has, in general, to vary
either the resonator natural frequency $\om$ or the external field
$B_0$, so that to achieve the approximate equality
$$
\gm_S B_0 \cong \om +(2S-1) \; \frac{D}{\hbar} \; s \qquad
(\Dlt_D=0 ) \; .
$$
In principle, such a temporal varying of the system parameters, for
achieving the resonance conditions, is feasible and is known for optical
systems [5], where it is called {\it chirping}. The chirping technique
can also be used for realizing superradiance from magnetic molecules
[60].

The situation becomes simple when $S=1/2$, so that the anisotropy
disappears, or when $\om_0\gg(2S-1)D/\hbar$. Then $\om_D\approx\om_0$,
and all the consideration reduces to the same as has been done above
for nuclear spins. Another possibility is by tuning the resonant
electric circuit to one of the transition frequencies of admissible
$2S$ transitions and by supporting the population of the upper level
with the help of the permanent pumping, as it was done for the nuclear
spin $I=5/2$ of $^{27}$Al in experiments [30--33]. In the latter case,
solely the regime of pulsing spin superradiance can be achieved.

In this way, though the presence of magnetic anisotropy complicates the
experimental realization of coherent spin radiation by magnetic molecules,
nevertheless there are several possibilities for reaching the required
conditions and producing such a coherent radiation. This could be used
for superradiant operation of spin masers [60]. Radiation from magnetic
molecules would be essentially stronger than that from nuclear spins.
And not only the current power (129) could be easily measured, but
radiation intensity (127) would also be very high. To estimate the
radiation intensity (127), let us take $\om_D\sim\om_0\sim 10^{12}$ Hz,
which corresponds to microwaves with the wavelength $\lbd\sim 0.1$ cm.
For the parameters typical for Mn$_{12}$ or Fe$_8$, we have $\Gm_2=n_0
\rho\gm_S\mu_S\sim 10^{10}$ s$^{-1}$ and $\Gm_0=0.1\Gm_2$. The
spin-resonator coupling (199) is quite large, $g_D\sim 10^5$. The
radiation intensity $\overline I\sim\mu_S^2\om_0^4N^2/c^3$, for
$N\sim 10^{20}$, reaches the value $\overline I\sim 10^{11}$ W. The
superradiant pulse can be very short. The pulse time $\tau_p\approx
T_2/gs_0$ can be as short as $10^{-15}$ s. So, it is feasible to
realize femtosecond superradiant pulses. The possibility of producing
coherent radiation by magnetic molecules can find various applications.

\section{Pion and Dibaryon Radiation}

Nuclear systems under extreme conditions can emit not only photons
but also other particles. If the emitted particles are bosons, they
can form coherent beams, similarly to the coherent photon radiation.
The required extreme conditions can be experimentally achieved in the
process of nuclear and heavy-ion collisions. Then for instance, multiple
pions can be produced. If the density of the latter is sufficiently high,
they can form a coherent state, thus, providing the feasibility of
getting a pion laser [112]. When the energy of colliding nuclei is high,
they can form fireballs of nuclear matter, having temperature and density
sufficient for realizing the deconfinement transition [113--118]. Then,
except pions, other bosons can be emitted, such as dibaryons and gluons,
presenting the possibility of obtaining different nuclear-matter lasers
[119].

In order to find out the conditions under which the massive creation of
bosons in nuclear matter could be achieved, it is necessary to analyze
the equation of state of hot and dense nuclear matter, taking into account
various channels of reactions where composite particles could be formed.
To our mind, the most general approach for reaching this goal is presented
by the {\it theory of clustering matter} [117,120,121]. This approach
is based on three pivotal concepts: cluster representation, statistical
correctness, and potential scaling.

The idea of {\it cluster representation} goes back to the works studying
the abundances of chemical elements by treating each element as a
quasiparticle characterized by its atomic weight and binding energy, with
the related chemical potentials taking account of the allowed chemical
reactions [122]. The problem of constructing an accurate quasiparticle
representation for composite particles was initiated by Weinberg [123--125].
The most mathematically elaborated approach was formulated by Girardeau
[126--128], whose method was applied to different systems containing bound
clusters, including quark-hadron matter [129,130].

The basic points in the cluster picture are as follows. Let us consider
a multiparticle system with the total space of quantum states being a Fock
space ${\cal F}$, on which the algebra of observables ${\cal A}$ is defined.
Let among these quantum states be free states as well as different bound
states corresponding to clusters of several particles. Each type of bound
clusters can be individualized by a set of characteristic parameters,
such as the compositeness number $z_i$, showing the number of elementary
particles bound into a cluster, effective mass of the cluster $m_i$, its
baryon number $B_i$, strangeness $S_i$, and so on. And let us treat each
type of these bound clusters as a separate sort of particles, with the
associated Fock space ${\cal F}_i$ called the ideal cluster space. The
{\it total cluster space} is defined as the tensor product
\be
\label{201}
{\cal F}_T \equiv \otimes_i \; {\cal F}_i \otimes {\cal F}
\ee
termed the {\it Fock-Tani space} [126--128,131]. The relation between the Fock
space
of elementary particles and the Fock-Tani cluster space (201) is
presented by means of the unitary Tani mapping $\hat U_T^+=\hat U_T^{-1}$,
so that
\be
\label{202}
{\cal F} = \hat U_T {\cal F}_T \; , \qquad
{\cal F}_T = \hat U_T^+{\cal F} \; .
\ee
The cluster algebra of observables is given by the Fock-Tani representation
\be
\label{203}
{\cal A}_T \equiv \hat U_T^+ {\cal A} \hat U_T \; .
\ee
The latter definition guarantees that all matrix elements of the algebra
${\cal A}$ on ${\cal F}$ are the same as those of ${\cal A}_T$ on ${\cal F}_T$,
since ${\cal F}_T^+{\cal A}_T{\cal F}_T={\cal F}^+{\cal A}{\cal F}$. As far
as the representation of ${\cal A}_T$ on ${\cal F}_T$ is isomorphic to that
of ${\cal A}$ on ${\cal F}$, all observable quantities are the same in the
standard picture of elementary particles and in the quasiparticle picture
of the cluster system.

The Tani mapping is constructed in the following way. Let $a_i(x)$ be the
field operator of elementary particles, defined on the Fock space ${\cal F}$,
with $x$ being a set of spatial variables. Suppose $\vp_p(x_1,x_2,\ldots,x_p)$
is a Schr\"odinger wave function describing a bound state of $p$ elementary
particles. The field operator of the associated bound cluster, defined on
the same space ${\cal F}$, can be presented as
$$
\Psi_p(x) \equiv \int \vp_p(x_1-x,x_2-x,\ldots,x_p-x)\; a_1(x_1)a_2(x_2)
\ldots a_p(x_p)\; dx_1 dx_2 \ldots dx_p \; .
$$
The image of this bound state on the ideal cluster space ${\cal F}_p$ is given
by a $p$-particle cluster described by the field operator $\psi_p(x)$. The Tani
mapping is defined by means of the unitary transformation
$$
\hat U_T \equiv \exp \left ( \frac{\pi}{2}\; \hat F\right ) \; ,
$$
$$
\hat F = \sum_p \int \left [ \psi^\dgr_p(x) \Psi_p(x) - \Psi_p^\dgr(x)
\psi_p(x) \right ] \; dx \; ,
$$ where the summation implies that over all admissible clusters. All operators
$a_i(x)$ and $\psi_p(x)$ satisfy the canonical commutation relations, depending
on whether the considered particle or cluster is a boson or fermion.

The algebra of observables (203) is defined on the Fock-Tani cluster space (201).
In particular, one has the cluster Hamiltonian
\be
\label{204}
H_T \equiv \hat U_T^+ H \hat U_T \; ,
\ee
where $H$ is the initial Hamiltonian in terms of elementary particles, given
on ${\cal F}$. The cluster statistical state is $<{\cal A}_T>$, with the angle
brackets implying the statistical averaging related to the cluster Hamiltonian
(204). The density of the $j$-type cluster is
$$
\rho_j =\frac{1}{V}\; <\hat N_j>\; , \qquad
\hat N_j \equiv \int \psi_j^\dgr (x) \psi_j(x) \; dx \; ,
$$
where $V$ is the system volume. We may introduce the {\it cluster probability}
\be
\label{205}
w_j \equiv z_j \; \frac{\rho_j}{\rho} \qquad
\left ( \rho\equiv \sum_j z_j \rho_j \right ) \; ,
\ee
characterizing the statistical weight of the $j$-type clusters. This probability
satisfies the standard properties, being semipositive and normalized,
$$
0\leq w_j \leq 1 \; , \qquad \sum_j w_j =1 \; .
$$

The direct construction of the cluster Hamiltonian (204) is a rather tedious
and complicated procedure. Moreover, the resulting Hamiltonian is presented
by an infinite series. Because of this it is customary to define an effective
Hamiltonian whose construction is based on physical reasoning. Generally,
an effective Hamiltonian $H_{eff}=H_{eff}(\{\rho_j\},T)$ includes an explicit
dependence on thermodynamic variables, such as cluster densities $\rho_j$
and temperature $T$. However, including thermodynamic variables into the
Hamiltonian may break the validity of the general thermodynamic relations.

The correct cluster Hamiltonian has to be defined so that to preserve
all thermodynamic relations. This is the meaning of the principle of
{\it statistical correctness} [117,120]. For this purpose, the effective
Hamiltonian is complimented by an additional term, guaranteeing the
statistical correctness of the resulting total cluster Hamiltonian
\be
\label{206}
H_T = H_{eff} + CV \; ,
\ee
which assumes the validity of the equations
\be
\label{207}
<\frac{\prt H_T}{\prt \rho_j}> \; = 0 \; , \qquad
<\frac{\prt H_T}{\prt T}> \; = 0 \; .
\ee
Choosing $C$ as a nonoperator quantity yields the equations
\be
\label{208}
\frac{\prt C}{\prt\rho_j}= -\; \frac{1}{V}\;
<\frac{\prt H_{eff}}{\prt\rho_j}> \; , \qquad \frac{\prt C}{\prt T} = -\;
\frac{1}{V}\; < \frac{\prt H_{eff}}{\prt T}> \; ,
\ee
defining $C=C(\{\rho_j\},T)$. These conditions guarantee the correctness of all
thermodynamic relations, such as
$$
P = -\; \frac{\prt\Om}{\prt V} = -\; \frac{\Om}{V} \; ,
$$
$$
\ep = T\; \frac{\prt P}{\prt T} - P + \mu_Bn_B + \mu_Sn_S = \frac{1}{V}\;
<H_T> \; , \quad s = \frac{\prt P}{\prt T} = \frac{1}{T}\;
\left (\ep + P -\mu_Bn_B -\mu_Sn_s\right ) \; ,
$$
\be
\label{209}
n_B = \frac{\prt P}{\prt\mu_B} = \sum_j B_j\rho_j \; , \qquad
n_S = \frac{\prt P}{\prt\mu_S} = \sum_j S_j\rho_j \; ,
\ee
in which $\Om$ is the grand potential, $\ep$ and $s$ are the energy and entropy
densities, $P$ is pressure, $\mu_B$ is baryon potential, $\mu_S$ is strangeness
potential, $n_B$ and $n_S$ are the baryon and strangeness densities, and
$$
\Om = - T\ln{\rm Tr} e^{-\bt\tilde H} \; , \qquad
\tilde H= H_T - \sum_j \mu_j N_j \; ,
$$
$$
\rho_j = \frac{\prt P}{\prt\mu_j} = \frac{1}{V}\; < \hat N_j> \; , \qquad
\mu_j = \mu_B B_j + \mu_S S_j \; ,
$$
with $\mu_j$ being the chemical potential and $\bt$, inverse temperature.

The cluster Hamiltonian (206) contains the terms describing effective
interactions between different clusters. Defining the related interaction
potentials is done by means of the principle of {\it potential scaling}
[117,120]. According to this, the interaction potentials from the same
class of universality are connected by the scaling relation
\be
\label{210}
\frac{\Phi_{ij}}{z_iz_j} = \frac{\Phi_{ab}}{z_az_b} \; .
\ee
This principle allows the definition of all qualitatively equivalent
interaction potentials through one known universal potential. Another
admissible form of the potential scaling could be
$$
\frac{\Phi_{ij}}{m_im_j} =\frac{\Phi_{ab}}{m_am_b} \; ,
$$
since masses are usually proportional to the corresponding compositeness
numbers, $m_j\sim z_j$.

The theory of clustering matter has been applied to clustering quark-hadron
matter, with elementary particles being quarks, antiquarks, and gluons,
while hadrons being various bound clusters of these elementary particles
[117,120].

The results of this consideration are presented in the following figures,
where the energy is measured in units of $J=225$ MeV, temperature
$\Theta=k_BT$ is given in MeV, $n_{0B}=0.167$ fm$^{-3}$ is the normal
baryon density. These results demonstrate that the deconfinement transition
at finite baryon density, and at conditions typical of heavy-ion collisions,
is a gradual crossover. The deconfinement transition can be associated with
a point where some reduced quantities display a maximum. The largest amount
of pions is produced around the deconfinement temperature $\Theta_d=160$ MeV
and
at the low baryon density $n_B<n_{0B}$. A high concentration of dibaryons
can be achieved at low temperature $\Theta<20$ MeV and the baryon density
$n_B\approx(5-20)n_{0B}$. Dibaryons can form the Bose-Einstein condensate,
thus, being in a coherent state. Producing high concentrations of pions
or dibaryons, under the corresponding conditions, can be employed for
creating
pion and dibaryon lasers [119]. For producing a large amount
of gluons, very
high temperatures, $\Theta>160$ MeV, are required. But gluons
cannot be emitted
as free particles.

\section{Conclusion}

We presented a detailed description of coherent radiation by nuclear
spins. It is important to stress that pure spin superradiance was first
experimentally observed in Dubna. The theoretical consideration of coherent
nuclear spin radiation is based on the theory developed by the authors.
The mathematical basis of this theory is the scale separation approach.

We show that at sufficiently strong external magnetic fields in the system
of nuclear spins there appear nuclear spin waves, which play the role of
the triggering mechanism starting the incoherent motion of spins. Neither
magnetodipole radiation nor the resonator Nyquist noise can initiate spin
motion. After the spins started moving, the transverse transition coherence
of their evolution is due to the action of the feedback magnetic field
formed by the resonant electric circuit. This resonator field is crucial
for developing coherent spin motion. Different regimes of coherent spin
radiation are investigated. Electron-nuclear hyperfine coupling can lead
to an essential enhancement of nuclear spin radiation. Superradiance by
magnetic molecules is feasible, which necessarily requires the presence
of a resonant electric circuit and cannot be achieved without it. The time
dependence of the transition frequency, caused by the single-site magnetic
anisotropy, can be compensated by the chirping effect. The possibility
of pion and dibaryon radiation from excited nuclear matter is discussed.

The mathematical techniques, presented in this review for treating strongly
nonequilibrium spin systems, are rather general and can be employed for
analysing nonlinear dynamics of arbitrary spin or quasispin essemblies. A
very interesting application of these techniques would be to investigating
the nonequilibrium phenomena occurring in dilute gases of cold trapped atoms
[95,132--135]. Such atoms can be spin-polarized, forming at low temperatures
the spinor Bose-Einstein condensates. The nonequilibrium dynamics of polarized
spinor condensates can exhibit coherent phenomena similar to those happening
in nuclear spin systems.

\vskip 1cm

{\bf Acknowledgements}

\vskip 2mm

This review summarizes the results of our 15-year work on the subject.
Throughout all these years we have enjoyed permanent support from
V.G. Kadyshevsky and A.N. Sissakian, for which we are so much grateful.
In the course of our work, we have discussed, at different times and in
different countries, many related problems with a number of people. We are
especially thankful for beneficial discussions, useful advice, and help
to A.M. Baldin, N.A. Bazhanov, F. Borsa, C.M. Bowden, M.G. Cottam,
B.C. Gerstein, B.N. Harmon, S.R. Hartmann, V.K. Henner, Y.F. Kisselev,
J.T. Manassah, P.P. Pashinin, M. Pruski, V.B. Priezzhev, V.V. Samartsev,
I.V. Yevseyev,
and V.M. Yermachenko.

\newpage

\begin{center}
{\large{\bf Figure Captions} }
\end{center}

{\bf Fig. 1}. Typical setup in experiments on spin superradiance. The left
coil plays the role of antenna. The right coil, surrounding the studied
sample, is part of the resonant electric circuit. The sample is shown as
a dark bar inside the resonant coil. The upper left block is an oscilloscope,
and the lower one is a plotter. The dashed line symbolizes refrigerator.

\vskip 3mm

{\bf Fig. 2}. Voltage signal of a superradiant pulse as a function of time,
measured in units of $10^{-7}$ s, for two initial spin polarizations:
$s_0=0.52$ (lower curve) and $s_0=0.57$ (upper curve).

\vskip 3mm

{\bf Fig. 3}. Intensity of radiation (upper curve) in arbitrary units and
the longitudinal spin polarization (lower curve) as functions of time,
measured in units of $T_2$, obtained from computer simulations for 300 spins.

\vskip 3mm

{\bf Fig. 4}. Orientation of the coordinate axes with respect to the spin
sample inserted into the coil of an electric circuit.

\vskip 3mm

{\bf Fig. 5}. Pressure (in units of $J^4$) of the clustering quark-hadron
matter.

\vskip 3mm

{\bf Fig. 6}. Energy density (in units of $J^4$) on the temperature-baryon
density plane.

\vskip 3mm

{\bf Fig. 7}. Pressure-to-energy density ratio defining the effective sound
velocity squared $c_{eff}^2\equiv P/\ep$, with a maximum at $\Theta_d=160$ MeV
related to the deconfinement.

\vskip 3mm

{\bf Fig. 8}. Specific heat $C_V=\prt\ep/\prt T$ (in units of $J^3$).

\vskip 3mm

{\bf Fig. 9}. Reduced specific heat $\sgm_V=TC_V/\ep$ displaying a maximum
at $\Theta_d=160$ MeV associated with the deconfinement crossover.

\vskip 3mm

{\bf Fig. 10}. Compression modulus $\kappa_T^{-1}=n_B\prt P/\prt n_B$
(in units of $J^4$) also having a maximum at the deconfinement crossover.

\vskip 3mm

{\bf Fig. 11}. Cluster probability of the quark-gluon plasma, being the
sum of the probabilities of quarks, antiquarks, and gluons.

\vskip 3mm

{\bf Fig. 12}. Pion probability, being the sum of the probabilities
of $\pi^+$, $\pi^-$, and $\pi^0$ mesons, with a sharp maximum at the
deconfinement crossover.

\vskip 3mm

{\bf Fig. 13}. Summary probability of $\eta$, $\rho^+$, $\rho^-$, $\rho^0$,
and $\om$ mesons.

\vskip 3mm

{\bf Fig. 14}. Nucleon probability, being the sum of the probabilities
of neutrons, protons, antineutrons, and antiprotons.

\vskip 3mm

{\bf Fig. 15}. Dibaryon probability, which is the sum of the dibaryon
and antidibaryon weights.

\vskip 3mm

{\bf Fig. 16}. Probability of the Bose-Einstein condensate of dibaryons.

\end{document}